\documentclass[pre,amsmath,aps,twocolumn,superscriptaddress,letterpaper]{revtex4-1}
\usepackage{graphicx,xspace,units,subfigure}
\usepackage{float}
\usepackage{amsmath}
\usepackage{amssymb}
\usepackage[normalem]{ulem}
\usepackage{appendix}
\input{epsf}

\newcommand{\noi}{\noindent}
\newcommand{\mbf}{\mathbf}
\newcommand{\be}{\begin{equation}}
\newcommand{\ee}{\end{equation}}

\begin{document}

%%%%%%%%%%%%%%%%%%%%%%%%%%%%%%%%%%%%%%%%%%%%%%%%%%%%%%%%%%%%%%%%%%%
\title{Correlations  in a Generalized Elastic Model: Fractional Langevin Equation Approach}

\author{Alessandro Taloni}
\affiliation{School of Chemistry, Tel Aviv University, Tel Aviv 69978, Israel}
\author{Aleksei Chechkin}
\affiliation{School of Chemistry, Tel Aviv University, Tel Aviv 69978, Israel}
\affiliation{Akhiezer Institute for Theoretical Physics, NSC KIPT,
  Kharkov 61108, Ukraine}
\author{Joseph Klafter}
\affiliation{School of Chemistry, Tel Aviv University, Tel Aviv 69978, Israel}

%%%%%%%%%%%%%%%%%%%%%%%%%%%%%%%%%%%%%%%%%%%%%%%%%%%%%%%%%%%%%%%%%%%
\begin{abstract}

The Generalized Elastic Model (GEM) provides the evolution equation which governs the stochastic motion  of several many-body systems in nature, such as polymers, membranes, growing interfaces. On the other hand  a probe (\emph{tracer}) particle in these systems performs a fractional Brownian motion due to the spatial interactions with the other system's components. The tracer's anomalous dynamics can be described by a Fractional Langevin Equation (FLE) with a space-time correlated noise. We demonstrate that the description given in terms of GEM coincides with that furnished by the relative FLE, by showing that the correlation functions of the stochastic field obtained within the FLE framework agree to the corresponding quantities calculated from the GEM. Furthermore we show that the Fox $H$-function formalism appears to be very convenient to describe the correlation properties within the FLE approach.

\end{abstract}
\maketitle

%%%%%%%%%%%%%%%%%%%%%%%%%%%%%%%%%%%%%%%%%%%%%%%%%%%%%%%%%%%%%%%%%%%
\section{Introduction}
\label{sec:Introduction}

Polymers~\cite{Doi,Granek,semiflexible}, elastic chains and
membranes ~\cite{Edwards,membranes,membranes-FLE,Granek,Zilman},
rough surfaces ~\cite{interfaces,Krug,surfaces}  can be described by
a continuum elastic model which accounts for their  general
stochastic behavior. This model, named Generalized Elastic Model
(GEM), is defined by the following stochastic linear
differential equation in partial derivatives
~\cite{Taloni-FLE},

\begin{equation}
\frac{\partial}{\partial t}\mathbf{h}\left(\vec{x},t\right)=\int d^dx'\Lambda\left(\vec{x}-\vec{x}'\right)\frac{\partial^z }{\partial\left|\vec{x}'\right|^z }\mathbf{h}(\vec{x}',t)+\boldsymbol\eta\left(\vec{x},t\right).
\label{GEM}
\end{equation}

\noi The stochastic field $\mathbf{h}$ is 
$D$-dimensional and defined in the $d-$dimensional infinite space. The white noise appearing in (\ref{GEM})
satisfies the fluctuation-dissipation (FD) relation, i.e.

\be
\langle\eta_{j}\left(\vec{x},t\right)\eta_{k}\left(\vec{x}',t'\right) \rangle
=  2k_BT\Lambda\left(\vec{x}-\vec{x}'\right)\delta_{j\,k}\delta(t-t').
\label{FDT}
\ee

\noi ($j,k\in[1,D]$) where $\Lambda\left(\vec{r}\right)$ corresponds to the hydrodynamic friction kernel
which couples different sites in $\vec{x}$ and $\vec{x'}$ through a fluid-mediated interaction.
The internal elastic coupling is instead embodied by the fractional derivative defined via its Fourier transform by ~\cite{Zazlawsky}

\be
{\cal F}_{\vec{q}}\left\{\frac{\partial^z}{\partial\left|\vec{x}\right|^z}\right\}\equiv-\left|\vec{q}\right|^z
\label{fractional_operator}.
\ee

\noi Another common definition is given in term of the Laplacian $\Delta$ as ~\cite{Samko}: $\frac{\partial^z}{\partial\left|\vec{x}\right|^z}:=-\left(-\Delta\right)^{z/2}$.

The systems whose the dynamical behavior is described by the GEM
(\ref{GEM}) can be divided into two different classes, according to
the type of hydrodynamic interactions that characterize them:
\emph{long ranged} or \emph{local}.

\subsection{Long ranged hydrodynamic interactions}

This situation occurs when the friction kernel is defined by a general expression like

\be
\Lambda\left(\vec{r}\right)\sim\frac{1}{\left|\vec{r}\right|^\alpha},
\label{hydro}
\ee

\noi where $a\ll\left|\vec{r}\right|\ll L$,    $L$ corresponds,
for example, to the screening length or the maximum size of the system,  and $a$ is the smallest length scale up to which the
continuum description furnished by (\ref{GEM}) keeps its validity.
Whenever our analysis will require a regularization at small and/or
long distances,  the largest and smallest length scales will then
come into play as the integral's upper and/or lower cut-offs.

\noi The hydrodynamic
interactions are often represented by
the equilibrium average of
the Oseen tensor, which in an embedding $d_e$-dimensional space ($d_e\geq
3$) reads, according to (\ref{hydro}):
$\Lambda\left(\vec{r}\right)\sim\left|\vec{r}\right|^{2-d_e}$  ~\cite{Doi,Zilman2}.
The following systems belong to this class.

- \emph{Fluid membranes}. The height of a fluctuating membrane is represented by the scalar quantity $h\left(\vec{x},t\right)$, i.e. $D=1$ ~\cite{membranes,membranes-FLE,Granek,Zilman}. The point $\vec{x}$ on the planar surface implies $d=2$, which in turn gives $d_e=D+d=3$ and $\alpha=1$. Since for small deformations ($\left|\vec{\nabla}h\right| \ll 1$) the bending free energy is $\propto \left(\Delta h\right)^2 $ ~\cite{Helfrich}, $z=4$ in (\ref{GEM}).

- \emph{Semiflexible polymers}. $\mbf{h}$ stands for the 3  spatial coordinates of a polymeric segment (\emph{bead}) while $x$ is
the strand's 1-dimensional internal coordinate (\emph{curvilinear
  abscissa}): $D=3$, $d=1$. The embedding dimension in this case coincides with $D$ ($d_e=3$), yielding $\alpha=1$ ~\cite{Granek,semiflexible} .
  The bending elastic energy associated with the chain's deformation implies  $z=4$ ~\cite{Harris} as in the previous example.

- \emph{Flexible polymers}. In this systems $\mbf{h}$ and $x$ still correspond to the  position and the curvilinear ascissa of the monomer in the polymeric chain: $D=3$, $d=1$. However the beads interaction is just represented by an harmonic coupling ($z=2$) and the Zimm's equilibrium approximation of the Oseen tensor in $\Theta$ solvent gives $\alpha=1/2$ ~\cite{Doi,Zimm}.

\subsection{Local hydrodynamic interactions}

This kind of systems present no fluid-mediated interactions, namely $\Lambda\left(\vec{r}\right)=\delta^d(\vec{r})$. This can be attributed either to the large screening among the elementary components of the system or to an interaction which is purely mechanical. Examples are:

- \emph{Rouse polymers}. Here, as in the case of (semi)flexible chains,
$\mbf{h}$ stands for the bead's 3-dimensional position, and $x$ for the bead's position along the chain ~\cite{Doi,Rouse}. The hookean interaction  gives $z=2$.

- \emph{Single file system}. The system of $N$ hard-core rods diffusing on a line without overlapping is known in literature as single file (see ~\cite{Lizana} and references therein). Recently it has been shown that the system's dynamics can be reduced, within a very good approximation, to a 1-dimensional harmonic chain problem (\emph{harmonization}) where $h(x,t)$ is the position of the $x$-th particle on the  substrate at time $t$.

- \emph{Fluctuating interfaces}. In this systems $h$ plays
the role of
a scalar field (mostly the height of a rough surface in $d$ dimension) which is
subjected to a
non-standard elastic force embodied by the fractional derivative of
order $z$ ~\cite{interfaces,Krug}. This is actually the generalization of the
Edwards-Wilkinson equation for the fluctuating profile of a granular
surface, for which $d=2$, $z=2$ ~\cite{Edwards}. In systems
such as crack propagation ~\cite{Gao} and contact line of a liquid
meniscus ~\cite{DeGennes} $d=1$, and the restoring forces are
characterized by $z=1$.

- \emph{Solid surfaces}. If  $h$ is a step, namely a line boundary at which   the surface changes height by one or more atomic units ~\cite{surfaces}, the value of $z$ in eq.(\ref{GEM}) is
found to be $z=2,3$ or $4$ ($d=1$)
according to the character of the
atomic diffusion: periphery, terrace or attachment-detachment diffusion respectively.

- \emph{Diffusion-noise equation}. In this case $h$
represents the density field
on a $d$-dimensional surface  $\vec{x}$ while the diffusion operator sets $z=2$ ~\cite{VKampen}.

\vspace{0.1cm}

The values of the parameters related to each of the models formerly
listed are summarized in table ~\ref{table}.

\vspace{0.5cm}

In ~\cite{Taloni-FLE} we addressed the question of the motion of a
tracer (\emph{probe}) particle in the systems whose dynamics
obeys to (\ref{GEM}). Although the whole system dynamics is
Markovian, the  particle placed at a position $\vec{x}$ undergoes a
subdiffusive motion on the score of persistent memory effects due to the spatial correlations with the rest of the system. Roughly speaking, a tracer
particle experiences two kinds of interactions: the first one is the
coupling with the surrounding heat bath,
whose mathematical expression is furnished by the Langevin
random force $\eta(\vec{x},t)$ and the corresponding FD relation
(\ref{FDT}). The second  interaction is inherent to the system: the
probe particle is coupled with the rest of the system  through 
both the hydrodynamic term (\ref{hydro}) and the fractional
Laplacian (\ref{fractional_operator}).  This
``internal''coupling  originates the spatial
correlations responsible for the tracer's long-ranged non-Markovian
memory effects. On the other hand, the non-Markovianity of the
probe particle's anomalous dynamics leads to the description in terms of fractional Brownian motion (FBM)
~\cite{Mandelbroot} which obeys a
fractional Langevin equation (FLE) ~\cite{Lizana,Lutz}. Within this
framework the strong internal interactions are mimicked by  the
colored noise term and the fractional derivative, connected by the
generalized fluctuation-dissipation relation.  In this article we
will show how the representation  of the tracer's stochastic
evolution given in terms of FLE offers the same level of accuracy
furnished by (\ref{GEM}). Indeed  any kind of physical statistical
observable can be evaluated starting from (\ref{GEM}) as well as
from the corresponding FLE.

This paper is outlined as follows: in Sec.\ref{sec:GEM_MSD} we start
from the GEM (\ref{GEM}) deriving the expressions for the
$h$-autocorrelation  function (and the corresponding mean square
displacement) in the case of  $z>d$,  $z<d$ and $z=d$
($(d-1)/2<\alpha  <  d$). In Section \ref{sec:FLE}, starting from
the expression (\ref{GEM}), we draw the FLE equation for the tracer
particle placed at position $\vec{x}$ when $z>d$. In Section
\ref{sec:noise} we derive the properties of the noise appearing in
the FLE and we demonstrate the validity of the
fluctuation-dissipation relation for the probe particle. Section
\ref{sec:h-autocorrelation} will be devoted  to the $h$-correlation
function arised from the FLE framework and we furnish its general
expression in terms of the Fox functions. In particular we
 apply  the developed
formalism to  systems such as fluid membranes,
proteins and fluctuating interfaces, recovering results previously derived in literature. In
Appendix \ref{app:hydro} we report the calculations for the
hydrodynamics term in the limiting case
$\alpha=d$. In Appendix \ref{app:FF} we  provide a
demonstration of the appeareance of the Fox functions in our
analysis, while  in Appendix \ref{app:FF_prop} are listed the
main properties of the Fox functions that we make use  of
through our calculations. Lastly, Appendix \ref{app:D-A_GLE} concerns with the derivation of the Generalized Langevin Equation for
the inter-monomeric distance in a 3-dimensional Rouse chain,
according to the procedure outlined in ~\cite{Lizana}.

\begin{table}[t]
\begin{tabular}{|c|c|c|c|c|}
    \hline
System                   & $D$       & $d$       & $z$        & $\alpha$    \\
    \hline
Fluid membranes ~\cite{membranes,membranes-FLE,Granek,Zilman}         & 1         & 2         & 4          &  1          \\
Semiflexible polymers ~\cite{Granek,semiflexible}   & 3         & 1         & 4          &  1           \\
Flexible polymers ~\cite{Doi,Zimm}       & 3         & 1         & 2          &  1/2         \\
Crack propagation   ~\cite{Gao}     & 1         & 1         & 1          &  -         \\
Liquid meniscus   ~\cite{DeGennes}        & 1         & 1         & 1          &  -            \\
Rouse polymers   ~\cite{Doi,Rouse}        & 3         & 1         & 2          &  -          \\
Single file systems  ~\cite{Lizana}    & 1         & 1         & 2          &  -         \\
Fluctuating interfaces ~\cite{interfaces,Krug,Edwards}  & 1         & any       & any        &  -         \\
Solid surfaces    ~\cite{surfaces}       & 1         & 1         & 2,3,4      &  -        \\
Diffusion-noise equation ~\cite{VKampen} & 1         & any       & 2          &  -         \\

        \hline
\end{tabular}
\caption{Values of the parameters $D,d,z$ and $\alpha$ charachterizing the GEM (\ref{GEM}) for the systems listed in the Introduction. Note that for systems whose the hydrodynamic interaction is local the value of $\alpha$ is absent.}
\label{table}
\end{table}

%%%%%%%%%%%%%%%%%%%%%%%%%%%%%%%%%%%%%%%%%%%%%%%%%%%%%%%%%%%%%%%%%%%%%%%%%%%%%%%
%
%         A U T O C O R R E L A T I O N  F U N C T I O N
%
%%%%%%%%%%%%%%%%%%%%%%%%%%%%%%%%%%%%%%%%%%%%%%%%%%%%%%%%%%%%%%%%%%%%%%%%%%%%%%%
\section{$h$-autocorrelation  function for the Generalized Elastic process}
\label{sec:GEM_MSD}

We start from the equation (\ref{GEM}). Defining Fourier transform of the
stochastic process in
space and time as

\be
\mathbf{h}\left(\vec{q},\omega\right)=\int_{-\infty}^{+\infty}d^dx \int_{-\infty}^{+\infty}dt\,
\mathbf{h}\left(\vec{x},t\right)\,e^{-i\left(\vec{q}\cdot\vec{x}-\omega t\right)},
\label{FFT}
\ee

\noi we find that the general solution of (\ref{GEM}) can be expressed
in the Fourier space in the following form

\be
\mathbf{h}\left(\vec{q},\omega\right)=\frac{\boldsymbol{\eta}\left(\vec{q},\omega\right)}{-i\omega+\Lambda\left(\vec{q}\right)\left|\vec{q}\right|^{z}},
\label{sol_FF}
\ee

\noi where $\Lambda\left(\vec{q}\right)$ is the Fourier
transform of hydrodynamic friction kernel,

\be
\Lambda\left(\vec{q}\right)\left\{
\begin{array}{lll}
=\frac{(4\pi)^{d/2}}{2^{\alpha}}\frac{\Gamma\left((d-\alpha)/2\right)}{\Gamma\left(\alpha/2\right)}\left|\vec{q}\right|^{\alpha-d}=A\left|\vec{q}\right|^{\alpha-d}
&   &  \frac{d-1}{2}<\alpha<d \\

\sim \frac{2\pi^{d/2}}{\Gamma\left(d/2\right)}\ln\left(\frac{1}{\left|\vec{q}\right|a}\right) &  & \alpha=d.
\end{array}
\right.
\label{hydro_FF}
\ee

\noi To get eq.(\ref{hydro_FF})  we use the expression for the $d$-dimensional Fourier transform of the isotropic function $\phi(\left|\vec{r}\right|)$ ~\cite{Champeney}

\be
\begin{array}{l}
\int_{-\infty}^{+\infty}d^dr e^{-i\vec{q}\cdot\vec{r}}\phi(\left| \vec{r}\right|)=\\
(2\pi)^{d/2}\left| \vec{q}\right|^{1-d/2}\int_{0}^{+\infty}d\left| \vec{r}\right| \left| \vec{r}\right|^{d/2}J_{d/2-1}(\left| \vec{q}\right|\left| \vec{r}\right|)\phi(\left| \vec{r}\right|),
\end{array}
\ee

\noi where $J_{d/2-1}$ represents the Bessel function of  fractional order $d/2-1$. The limiting case $\alpha=d$ requires a regularization of
the integral in the $\vec{q}$-space: this has been done introducing
the cut-off $a$. Note that if we resort to the different definition
$\Lambda\left(\vec{r}\right)\sim\frac{1}{a^d+\left|\vec{r}\right|^d}$,
we get the same asymptotic expansion (for small $\left|\vec{q}\right|$) obtained in (\ref{hydro_FF}) for $\alpha=d$;
in this case however, the complete results for $d=1,\,2 $ and 3 are
presented in Appendix \ref{app:hydro}. The case $\alpha=(d-1)/2$
which requires the infrared cut-off $L$ won't be treated in the
following. 

\noi In the local
hydrodynamic situation $\Lambda\left(\vec{r}\right)=\delta\left(\vec{r}\right)$, which corresponds to take $\Lambda\left(\vec{q}\right)=1$ in (\ref{sol_FF}). Therefore  the calculations for local hydrodynamic systems can be either performed starting from (\ref{sol_FF}) where $\Lambda\left(\vec{q}\right)=1$, or simply  
setting $A=const$
and $\alpha=d$ in the corresponding long-ranged hydrodynamic expressions: this
substitution, \emph{which is not to be intended as a limit},  allows 
to easily shift from power-law to local hydrodynamics throughout the
following analysis. Indeed it corresponds to a formal  procedure  to   obtain the results for local hydrodynamic systems, starting from the equivalent quantities elaborated for long ranged hydrodynamic models. As a consequence, the case $\alpha=d$ and the ensuing logarithmic behavior (\ref{hydro_FF}), does not have to be confused with the systems with local hydrodynamic interactions. 

\noi We then define the $h$-autocorrelation function as $\langle
\delta h\left(\vec{x},t\right)\delta h\left(\vec{x},t'\right)
\rangle = \langle\left[
h_{j}\left(\vec{x},t\right)-h_{j}\left(\vec{x},0\right)\right]\left[h_{j}\left(\vec{x},t'\right)-h_{j}\left(\vec{x},0\right)
\right]\rangle$. Because of the isotropy of the system henceforth we
will drop the index $j$.  Since the Fourier transform of the noise
correlation function (\ref{FDT}) gets the form
$\langle\eta_j\left(\vec{q},\omega\right)\eta_k\left(\vec{q}',\omega'\right)
\rangle =2 k_BT
(2\pi)^{d+1}\Lambda\left(\vec{q}\right)\delta_{j\,k}\delta\left(\omega+\omega'\right)\delta^d\left(\vec{q}+\vec{q}'\right)$,
after a bit of algebra we derive

\be
\begin{array}{l}
\langle
\delta h\left(\vec{x},t\right)\delta h\left(\vec{x},t'\right)
\rangle  = k_BT\times\\
\times \int_{-\infty}^{+\infty}\frac{d^dq}{(2\pi)^d}\frac{g\left(\left|\vec{q}\right|,t\right)+g\left(\left|\vec{q}\right|,t'\right)-g\left(\left|\vec{q}\right|,\left|t-t'\right|\right)}{\left|\vec{q}\right|^{z}}\\
\begin{split}
g\left(\left|\vec{q}\right|,t\right)=1-e^{-\Lambda\left(\vec{q}\right)\left|\vec{q}\right|^{z}t}.
\end{split}
\end{array}
\label{EW_corr_func}
\ee

\noi When $(d-1)/2<\alpha<d$ the general expression of (\ref{EW_corr_func})
has the following form:

\be
\langle
\delta h\left(\vec{x},t\right) \delta h\left(\vec{x},t'\right)
\rangle =
K\left[f(t)+f\left(t'\right)-f\left(\left|t-t'\right|\right)\right].
\label{EW_corr_func_general_form}
\ee

\noi Three different situations can occur.

\emph{i)} $z<d$  ($\alpha>d-z$). The integrals in (\ref{EW_corr_func}) are
divergents when $\left|\vec{q}\right|\to\infty$: hence once again we are  compelled
to introduce the cut-off $a$. Performing the integrations yields

\be
\begin{array}{l}
K  =  \frac{2k_BT\pi^{d/2}}{(2\pi)^d\Gamma\left(d/2\right)}\frac{A^{\beta}}{\alpha+z-d}\\
f(t)=
-\beta\left[A\left(\frac{\pi}{a}\right)^{\alpha+z-d}\right]^{-\beta}-\gamma\left(-\beta,A\left(\frac{\pi}{a}\right)^{\alpha+z-d}t\right)t^{\beta},
\label{EW_corr_func_z<d}
\end{array}
\ee

\noi where 

\be
\beta=\frac{z-d}{\alpha+z-d}
\label{beta}
\ee

\noi and the function
$\gamma(a,x)$ is defined in
~\cite{Abramowitz}. It
is straightforward to verify that in this case the mean square
displacement gets to a constant value, $\langle
\delta^2 h\left(\vec{x},t\right)\rangle\to
\frac{4k_BT\pi^{d/2}}{(2\pi)^d\Gamma\left(d/2\right)(d-z)}\left(\frac{\pi}{a}\right)^{d-z}$. Physically this means that the system is overconnected and asymptotically any probe remains trapped around its initial position.

 \emph{ii)} $z=d$.  The integrals in (\ref{EW_corr_func}) exhibit
 a logarithmic divergence. After the regularization through the
 insertion of $a$ the result is
 achieved substituting in
(\ref{EW_corr_func_general_form})

\be
\begin{array}{l}
K  =  \frac{2k_BT\pi^{d/2}}{(2\pi)^d\Gamma\left(d/2\right)\alpha}\\
f(t)=
E_1\left(A\left(\frac{\pi}{a}\right)^{\alpha}t\right)+\ln\left[A\left(\frac{\pi}{a}\right)^{\alpha}t\right]+C,
\label{EW_corr_func_z=d}
\end{array}
\ee

\noi where $C$ is a constant value that we set to 0 and
 $E_1(x)$ denotes the exponential integral ~\cite{Abramowitz}. In
this case the system attains an asymptotic logarithmic
diffusion: $\langle\delta^2 h\left(\vec{x},t\right)\rangle\to
\frac{4k_BT\pi^{d/2}}{(2\pi)^d\Gamma\left(d/2\right)\alpha}\ln\left[A\left(\frac{\pi}{a}\right)^{\alpha}t\right]$. This is a borderline case, where the probe is not completely free to diffuse away from its initial position: the ensuing erratic motion is then logarithmic.

\emph{iii)} $z>d$. This is the most interesting situation: in this case the integrals can be performed effortlessly to give

\be
\begin{array}{l}
K  =  \frac{2k_BT\pi^{d/2}}{(2\pi)^d\Gamma\left(d/2\right)}\frac{A^{\beta}\Gamma\left(1-\beta\right)}{z-d}\\
f(t)= t^{\beta}.
\label{EW_corr_func_z>d}
\end{array}
\ee

\noi Therefore, the tracer particle placed at a given $\vec{x}$ performs
a subdiffusive fractional Brownian motion (FBM) ~\cite{Mandelbroot} given by  $\langle\delta^2 h\left(\vec{x},t\right)\rangle=2Kt^{\beta}$. In the language of fluctuating interfaces an interface is called \emph{rough} in this case ~\cite{Krug}.

In the following Sections we will focus on  the situation $iii)$
for which  $z>d$.

%%%%%%%%%%%%%%%%%%%%%%%%%%%%%%%%%%%%%%%%%%%%%%%%%%%%%%%%%%%%%%%%%%%%%%%%%%%%%%%
%
%                                  F L E
%
%%%%%%%%%%%%%%%%%%%%%%%%%%%%%%%%%%%%%%%%%%%%%%%%%%%%%%%%%%%%%%%%%%%%%%%%%%%%%%%
\section{Fractional Langevin Equation}
\label{sec:FLE}

Starting from the solution (\ref{sol_FF}) we now discuss the
derivation of the fractional Langevin equation for the probe particle
placed at a given position $\vec{x}$.
%Defining the constant $K^+$ as

%\be
%K^+= \frac{(2\pi)^{d-1}\Gamma(d/2)}{\pi^{d/2}A^\beta}(z+\alpha-d)\sin\left[(1-\beta)\pi\right]
%\label{K+}
%\ee

\noi Let us first multiply both sides of (\ref{sol_FF}) by
$K^+(-i\omega)^{\beta}$. As one can see from the derivation presented below, this is the only choice leading to the FLE which obeys the FD relation. According to ~\cite{Taloni-FLE}, choosing another arbitrary power instead of $(-i\omega)^{\beta}$ would lead to another equation among the family of Generalized Fractional Langevin Equations (GFLE), but without FD relation fulfiled. Thus eq.(\ref{sol_FF}) becomes

\be
K^+(-i\omega)^{\beta}h\left(\vec{q},\omega\right)=\frac{K^+(-i\omega)^{\beta}}{-i\omega+A\left|\vec{q}\right|^{z+\alpha-d}}\,\eta\left(\vec{q},\omega\right).
\label{sol_FF_1}
\ee

\noi $K^+$ is a constant that we introduce in order to fulfill the
fluctuation-dissipation relation: indeed this physical constraint
must be satisfied regardless of the description made  of the tracer's dynamics, i.e. both by the Markovian Langevin description given in
(\ref{GEM}) and by the fractional Langevin representation that we
are aimed at deriving. 

\noi Equation(\ref{sol_FF_1}) can be rewritten as \be
K^+(-i\omega)^{\beta}h\left(\vec{q},\omega\right)=K^+\eta\left(\vec{q},\omega\right)\Phi\left(\vec{q},\omega\right)
, \label{rewrite sol FF 1} \ee
where we introduce the function
$\Phi\left(\vec{x},t\right)$ whose Fourier transform in space and
time reads 

\be
\Phi\left(\vec{q},\omega\right)=\frac{(-i\omega)^{\beta}}{-i\omega+A\left|\vec{q}\right|^{z+\alpha-d}},
\label{Phi_FF} \ee

\noi Now, we simply make an inverse Fourier transform of Eq.(\ref{rewrite
sol FF 1}) in space and time. In the right hand side we get a new
noise term $\zeta\left(\vec{x},t\right)$, which is the convolution of
$\Phi\left(\vec{x},t\right)$ with the white Gaussian noise
$\eta\left(\vec{x},t\right)$, i.e.

\be
\zeta\left(\vec{x},t\right)=K^+\int_{-\infty}^{\infty}
d^dx'\int_{-\infty}^{\infty}
dt'\,\eta\left(\vec{x}-\vec{x}',t-t'\right)\Phi\left(\vec{x}',t'\right).
\label{FGN}
\ee

\noi To transform the left hand side of Eq.(\ref{rewrite sol FF 1}),
we introduce the Caputo derivative with lower bound equal to
$-\infty$, which for a "sufficiently well-behaved" function
$\phi(t)$ is defined as follows ~\cite{Samko,Podlubny},

\be
D_+^{\beta}\phi(t)=\frac{1}{\Gamma\left(1-\beta\right)}\int_{-\infty}^tdt'\frac{1}{\left(t-t'\right)^{\beta}}\frac{d}{dt'}\phi\left(t'\right),
\     \ 0<\beta<1, \label{RL} \ee

\noi  and whose Fourier transform reads as

\be \int_{-\infty}^{\infty}dt e^{i\omega
t}D_+^{\beta}\phi(t)=(-i\omega)^{\beta}\phi(\omega) \label{RL_FF}.
\ee

 \noi Thus, we get finally the fractional Langevin equation for the stochastic field $h\left(\vec{x},t\right)$
~\cite{membranes-FLE,Lutz,Lizana,Kou,Taloni-FLE},

\be
K^+D_+^{\beta}h\left(\vec{x},t\right)=\zeta\left(\vec{x},t\right).
\label{FLE} 
\ee

\noi From Eq.(\ref{FLE}) and the definition (\ref{RL}), the requirement of the validity of the
fluctuation-dissipation relation reads

\be
\langle \zeta\left(\vec{x},t\right)
\zeta\left(\vec{x},t'\right)\rangle=k_BT \frac{K^+}{\Gamma\left(1-\beta\right)\left|t-t'\right|^{\beta}},
\label{FLE-FDT}
\ee

\noi which relates the autocorrelation function of the noise ( standing in the right hand side of eq.(\ref{FLE})) to the damping kernel (determined by the fractional derivative (\ref{RL})) in the left hand side of eq.(\ref{FLE}). From this requirement the value of $K^+$ will be set. This will be done in
the next Section. It is important to remark that the random field (\ref{FGN}) is still Gaussian, since it is the linear combination of the Gaussian noise $\eta(\vec{x},t)$ but is also power-law correlated in time according to the FD relation (\ref{FLE-FDT}), hence is a fractional Gaussian noise (fGn).

%%%%%%%%%%%%%%%%%%%%%%%%%%%%%%%%%%%%%%%%%%%%%%%%%%%%%%%%%%%%%%%%%%%%%%%%%%%%%%%
%
%                N O I S E C O R R E L A T I O N
%
%%%%%%%%%%%%%%%%%%%%%%%%%%%%%%%%%%%%%%%%%%%%%%%%%%%%%%%%%%%%%%%%%%%%%%%%%%%%%%%
\section{Fractional Gaussian Noise correlation function}
\label{sec:noise}

We want to evaluate the two-point two-time correlation function of the
fGn      appearing in (\ref{FLE}). From (\ref{FDT}) and from the
definition (\ref{FGN}) we obtain

\be
\begin{array}{l}
\langle \zeta\left(\vec{x},t\right)
\zeta\left(\vec{x'},t'\right)\rangle=2k_BT
K^{+2}\times\\
\begin{split}
\times\int_{-\infty}^{+\infty}\frac{d\omega}{2\pi}e^{-i\omega\left(t-t'\right)}\varphi\left(\vec{x}-\vec{x'},\omega\right)
\end{split}
\label{FGN_cc_1}
\end{array}
\ee

\noi where

\be
\begin{array}{l}
\varphi\left(\vec{x}-\vec{x'},\omega\right)=\\
\int
d\vec{x_1}d\vec{x_2}\Lambda\left(\vec{x_1}-\vec{x_2}\right)\Phi\left(\vec{x}-\vec{x_1},\omega\right)\Phi\left(\vec{x'}-\vec{x_2},-\omega\right).
\label{phi}
\end{array}
\ee

\noi To calculate (\ref{phi}) we make use of the hydrodynamic term's
Fourier transform (\ref{hydro_FF}) and of (\ref{Phi_FF}): after
straightforward manipulations it takes the form

\be
\begin{array}{l}
\varphi\left(\vec{x}-\vec{x'},\omega\right)=\frac{A}{(2\pi)^{d/2}}|\omega|^{2\beta}\left|\vec{x}-\vec{x'}\right|^{1-d/2}\times\\
\times\int_{0}^{\infty}d\left|\vec{q}\right|\frac{\left|\vec{q}\right|^{\alpha-d/2}J_{d/2-1}\left(\left|\vec{q}\right|\left|\vec{x}-\vec{x'}\right|\right)}{\omega^2+A^2\left|\vec{q}\right|^{2(z+\alpha-d)}},
\label{phi_1}
\end{array}
\ee

\noi which, after a change of variable, becomes

\be
\begin{array}{l}
\varphi\left(\vec{x}-\vec{x'},\omega\right)=\frac{A^{\frac{2z-d-2}{2(z+\alpha-d)}}}{(2\pi)^{d/2}}|\omega|^{\frac{-2\alpha-d+2}{2(z+\alpha-d)}}\left|\vec{x}-\vec{x'}\right|^{1-d/2}\times\\
\times\int_{0}^{\infty}dy \frac{y^{\alpha-d/2}}{1+y^{2(z+\alpha-d)}}J_{d/2-1}\left(\frac{y|\omega|^{1/(z+\alpha-d)}\left|\vec{x}-\vec{x'}\right|}{A^{1/(z+\alpha-d)}}\right)
\label{phi_2}
\end{array}
\ee

\noi To proceed further, at this stage we employ the formalism of the Fox $H-$functions. These functions, introduced by Fox in 1961 ~\cite{Fox}, are special functions of a very general nature which allow us to present the results in a universal and elegant fashion. For a general theory on the $H-$functions we address the reader to the monograph of Mathai and Saxena ~\cite{Mathai} and to ref.~\cite{Hilfer}. As interesting applications of  $H-$functions we could mention an exactly solvable model of linear viscoelastic behavior ~\cite{Nonnenmacher1}, the $H-$function representation of non-Debye relaxation ~\cite{Nonnenmacher2,Hilfer2} and of the solution of the space-time fractional diffusion equations ~\cite{Metzler,Mainardi}. We present the defintion and the basic properties of the $H-$functions in Appendices \ref{app:FF} and \ref{app:FF_prop}, respectively.  Moreover in Appendix \ref{app:FF} it is shown that the function $1/(1+y^{2(z+\alpha-d)})$
appearing in the expression (\ref{phi_2}) can be cast in term of a Fox function:

\be
\frac{1}{1+y^{\gamma}}=\frac{1}{\gamma}H_{1\,1}^{1\,1}\left[y\left|{\begin{array}{c}
\left(0,\frac{1}{\gamma}\right) \\
\left(0,\frac{1}{\gamma}\right)\\
\end{array} } \right.\right]
\label{FF_appearence}
\ee

\noi where we introduced the short notation 

\be
\gamma=2(z+\alpha-d)
\label{gamma}
\ee

\noi After this substitution (\ref{phi_2}) reads

\be
\begin{array}{l}
\varphi\left(\vec{x}-\vec{x'},\omega\right)=\frac{A^{\beta+\frac{d-2}{\gamma}}}{(2\pi)^{d/2}\gamma}|\omega|^{\frac{-2\alpha-d+2}{\gamma}}\left|\vec{x}-\vec{x'}\right|^{1-d/2}\times\\
\times\int_{0}^{\infty}dy \, y^{\alpha-d/2}J_{d/2-1}\left(ky\right)H_{1\,1}^{1\,1}\left[y\left|{\begin{array}{c}
\left(0,\frac{1}{\gamma}\right) \\
\left(0,\frac{1}{\gamma}\right)\\
\end{array} } \right.\right]
\label{phi_3}
\end{array}
\ee

\noi with
$k=(|\omega|/A)^{2/\gamma}\left|\vec{x}-\vec{x'}\right|$.
Using the property (\ref{pr.5}) the integral can be evaluated to give

\be
\begin{array}{l}
\varphi\left(\vec{x}-\vec{x'},\omega\right)=\frac{A^{\beta+\frac{d-2}{\gamma}}}{(2\pi)^{d/2}\gamma}|\omega|^{\frac{-2\alpha-d+2}{\gamma}}\left|\vec{x}-\vec{x'}\right|^{1-d/2}\times\\
\frac{2^{\alpha-d/2}}{k^{\alpha+1-d/2}}
H_{3\,1}^{1\,2}\left[\frac{2}{k}\left|{\begin{array}{ccc}
\left(1-\frac{\alpha}{2},\frac{1}{2}\right)&\left(0,\frac{1}{\gamma}\right)&\left(1-\frac{\alpha+2-d}{2},\frac{1}{2}\right) \\
 &\left(0,\frac{1}{\gamma}\right)& \\
\end{array} } \right.\right],
\label{phi_4}
\end{array}
\ee

\noi  which, thanks to (\ref{1.2.3}) and
(\ref{1.2.5}), gets the form

\be
\begin{array}{l}
\varphi\left(\vec{x}-\vec{x'},\omega\right)=\frac{A^{\beta+\frac{d-2}{\gamma}}}{2(2\pi)^{d/2}\gamma}|\omega|^{\frac{-2\alpha-d+2}{\gamma}}\left|\vec{x}-\vec{x'}\right|^{1-d/2}\times\\
H_{1\,3}^{2\,1}\left[\frac{k}{2}\left|{\begin{array}{ccc}
&\left(\frac{4z+2\alpha-3d-2}{2\gamma},\frac{1}{\gamma}\right)& \\
\left(\frac{d-2}{4},\frac{1}{2}\right)&\left(\frac{4z+2\alpha-3d-2}{2\gamma},\frac{1}{\gamma}\right)&\left(\frac{2-d}{4},\frac{1}{2}\right)\\
\end{array} } \right.\right].
\label{phi_5}
\end{array}
\ee

\noi Applying again the property (\ref{1.2.5}), the noise
correlation function (\ref{FGN_cc_1}) can be written as

\be
\begin{array}{l}
\langle \zeta\left(\vec{x},t\right)
\zeta\left(\vec{x'},t'\right)\rangle=\frac{2k_BT
K^{+2}A^{2\beta-1}}{2^{\alpha+d}\gamma\pi^{d/2+1}}\left|\vec{x}-\vec{x'}\right|^{\alpha}\int_{0}^{+\infty}d\omega\times\\
\times
\cos\left(\omega\left(t-t'\right)\right)H_{1\,3}^{2\,1}\left[\frac{k}{2}\left|{\begin{array}{ccc}
&\left(\beta,\frac{1}{\gamma}\right)& \\
\left(\frac{-\alpha}{2},\frac{1}{2}\right)&\left(\beta,\frac{1}{\gamma}\right)&\left(\frac{2-\alpha-d}{2},\frac{1}{2}\right)\\
\end{array} } \right.\right],
\end{array}
\label{FGN_cc_2}
\ee

\noi where $\beta$  and $\gamma$ are given by (\ref{beta}) and
(\ref{gamma}) respectively.  The integral in (\ref{FGN_cc_2}) can be solved by referring  to the
property (\ref{pr.4}) of the Fox function: the final expression for
the noise correlation function then reads

\be
\begin{array}{l}
\langle \zeta\left(\vec{x},t\right)
\zeta\left(\vec{x'},t'\right)\rangle=\frac{k_BT
K^{+2}A^{2\beta-1}}{2^{\alpha+d}\pi^{(d+1)/2}}\frac{\left|\vec{x}-\vec{x'}\right|^{\alpha}}{\left|t-t'\right|}\times\\
\times
H_{3\,3}^{2\,2}\left[\frac{2}{A\left|t-t'\right|}\left(\frac{\left|\vec{x}-\vec{x'}\right|}{2}\right)^{\frac{\gamma}{2}}\left|{\begin{array}{ccc}
\left(\frac{1}{2},\frac{1}{2}\right)&\left(\beta,\frac{1}{2}\right)& \left(0,\frac{1}{2}\right)\\
\left(-\frac{\alpha}{2},\frac{\gamma}{4}\right)&\left(\beta,\frac{1}{2}\right)&\left(\frac{2-\alpha-d}{2},\frac{\gamma}{4}\right)\\
\end{array} } \right.\right].
\end{array}
\label{FGN_cc_3}
\ee

\noi The former expression is the central result of this paper. It states that the fGn entering the FLE (\ref{FLE}) is not only correlated in time, as required by the physical constraint (\ref{FLE-FDT}), but it is also space-correlated. This means that the space correlations  appearing in (\ref{GEM}), which are embodied by the hydrodynamic term as well as by the fractional Laplacian, are translated into space-time correlations of the noise in the FLE dynamical representation.

We recall now that the coefficient $K^+$ in (\ref{FGN_cc_3}) and (\ref{FLE-FDT}) is still undefined, our aim is to set its expression.  For this purpose we calculate the autocorrelation function of the
noise, i.e. we set  $\vec{x}\equiv\vec{x'}$ in
(\ref{FGN_cc_3}). To do so,  we first
expand the Fox function for small argument according to (\ref{exp})
and restric ourselves to the main term which diverges at
$\vec{x}\equiv\vec{x`}$:

\be
\begin{array}{l}
H_{3\,3}^{2\,2}\left[\frac{2}{A\left|t-t'\right|}\left(\frac{\left|\vec{x}-\vec{x'}\right|}{2}\right)^{\frac{\gamma}{2}}\left|{\begin{array}{ccc}
\left(\frac{1}{2},\frac{1}{2}\right)&\left(\beta,\frac{1}{2}\right)& \left(0,\frac{1}{2}\right)\\
\left(-\frac{\alpha}{2},\frac{\gamma}{4}\right)&\left(\beta,\frac{1}{2}\right)&\left(\frac{2-\alpha-d}{2},\frac{\gamma}{4}\right)\\
\end{array} } \right.\right]=\\
\frac{2^{2+\alpha}\sqrt{\pi}A^{1-\beta}}{\gamma}\frac{\Gamma(\beta)}{\Gamma(d/2)}\frac{\left|t-t'\right|^{1-\beta}}{\left|\vec{x}-\vec{x'}\right|^{\alpha}}.
\end{array}
\label{FGN_cc_exp_1}
\ee

\noi We then plug such expression in (\ref{FGN_cc_3}), achieving the
following final form for the fractional Gaussian noise autocorrelation
function :

\be
\langle \zeta\left(\vec{x},t\right)
\zeta\left(\vec{x},t'\right)\rangle=\frac{k_BT}{\left|t-t'\right|^{\beta}}
K^{+2}\frac{A^{\beta}2^{2-d}}{\pi^{d/2}\gamma}\frac{\Gamma(\beta)}{\Gamma(d/2)}.
\label{FGN_cc_exp_2}
\ee

\noi Thus the comparison between (\ref{FGN_cc_exp_2}) and
(\ref{FLE-FDT}) yields the value of  $K^+$, namely

\be
K^+=(4\pi)^{d/2-1}\gamma\sin\left(\pi\beta\right)\frac{\Gamma(d/2)}{A^\beta}.
\label{K+}
\ee

%%%%%%%%%%%%%%%%%%%%%%%%%%%%%%%%%%%%%%%%%%%%%%%%%%%%%%%%%%%%%%%%%%%%%%%%%%%%%%%
%
%                H C O R R E L A T I O N
%
%%%%%%%%%%%%%%%%%%%%%%%%%%%%%%%%%%%%%%%%%%%%%%%%%%%%%%%%%%%%%%%%%%%%%%%%%%%%%%%
\section{Two-point two-time $h$-correlation function}
\label{sec:h-autocorrelation}

In this Section we address the problem of the evaluation of the
$h$-correlation function within the framework of the FLE. Indeed, any
kind of statistical  observable can be expressed in term of
correlation functions, whose analytical expression can be derived either
starting from the eq.(\ref{GEM}) or from eq.(\ref{FLE}). Although the
representation of the system dynamics can be different, the
correlation functions must coincide, since the observable physical properties do
not have to depend on the chosen description. Therefore we can
furnish the general expression of the correlation function starting
from the solution of (\ref{FLE}).

\noi The solution of (\ref{FLE}) in the Fourier-Fourier space reads

\be
h\left(\vec{q},\omega\right)=\frac{\zeta\left(\vec{q},\omega\right)}{K^+(-i\omega)^{\beta}},
\label{sol_FLE_FF}
\ee

\noi and consequently the two-point two-time correlation function  is

\be
\begin{array}{l}
\langle h\left(\vec{x},t\right)
h\left(\vec{x'},t'\right)\rangle=\\
\begin{split}
\frac{1}{K^{+2}}\int\frac{d\omega\,d\omega'}{4\pi^2}
  e^{-i\omega t}e^{-i\omega' t'}\frac{\langle \zeta\left(\vec{x},\omega\right)
\zeta\left(\vec{x'},\omega'\right)\rangle}{(-i\omega)^{\beta}(-i\omega')^{\beta}}
\end{split}
\label{h_corr_FLE}.
\end{array}
\ee

\noi Using (\ref{FGN_cc_2}), the Fox function expression for
(\ref{h_corr_FLE}) reads

\be
\begin{array}{l}
\langle h\left(\vec{x},t\right)
h\left(\vec{x'},t'\right)\rangle=\frac{2k_BT
A^{2\beta-1}}{2^{\alpha+d}\gamma\pi^{d/2+1}}\left|\vec{x}-\vec{x'}\right|^{\alpha}\int_{0}^{+\infty}d\omega
\times\\
\times\frac{\cos\left(\omega\left(t-t'\right)\right)}{\omega^{2\beta}}
H_{1\,3}^{2\,1}\left[\frac{k}{2}\left|{\begin{array}{ccc}
&\left(\beta,\frac{1}{\gamma}\right)& \\
\left(\frac{-\alpha}{2},\frac{1}{2}\right)&\left(\beta,\frac{1}{\gamma}\right)&\left(\frac{2-\alpha-d}{2},\frac{1}{2}\right)\\
\end{array} } \right.\right].
\end{array}
\label{h_corr_FLE_2}
\ee

\noi The previous expression constitutes the elementary component
starting with whom,
any kind of physical observable is constructed. Nonetheless,  the
integral appearing in (\ref{h_corr_FLE_2}) cannot be solved as we performed in (\ref{FGN_cc_2}). The reason is that such integral is
divergent in the limit $\omega\to 0$, as it is apparent by an expansion of the
Fox function at small argument, i.e., recalling that $k=\left(\frac{\left|\omega\right|}{A}\right)^{2/\gamma}\left|\vec{x}-\vec{x'}\right|$ and using
(\ref{exp}), one has

\be
\begin{array}{l}
H_{1\,3}^{2\,1}\left[\frac{k}{2}\left|{\begin{array}{ccc}
&\left(\beta,\frac{1}{\gamma}\right)& \\
\left(\frac{-\alpha}{2},\frac{1}{2}\right)&\left(\beta,\frac{1}{\gamma}\right)&\left(\frac{2-\alpha-d}{2},\frac{1}{2}\right)\\
\end{array} } \right.\right]\simeq\\
\frac{2\pi}{\Gamma(d/2)\cos\left(\beta/2\right)}\left(\frac{2}{\left|\vec{x}-\vec{x'}\right|}\right)^{\alpha}\left(\frac{A}{\left|\omega\right|}\right)^{1-\beta}.
\label{FF_H_exp}
\end{array}
\ee

\noi  It follows from (\ref{FF_H_exp}) and (\ref{h_corr_FLE_2}) that the integrand in (\ref{h_corr_FLE_2}) diverges as $\sim
\left|\omega\right|^{-(1+\beta)}$. Therefore,  in order be
measurable, any
physical observable has to be arranged in such a way that the
divergence will be eliminated. In the following subsections we will
analyze some specific case of statistical quantities constructed as
linear combination of (\ref{h_corr_FLE_2})  and compare them with the corresponding
quantities arising from (\ref{sol_FF}).

\subsection{Mean Square Displacement}

The mean square displacement of the probe particle has been defined
in secton \ref{sec:GEM_MSD} as the limit $t\to t'$ of the
autocorrelation function (\ref{EW_corr_func}); in term of elementary component
(\ref{h_corr_FLE_2}) it can be expressed as

\be
\langle
\delta^2 h\left(\vec{x},t\right)\rangle  = 2\left[ \langle
h^2\left(\vec{x},0\right)\rangle -\langle h\left(\vec{x},t\right) h\left(\vec{x},0\right)\rangle\right].
\label{FLE_MSD}
\ee

\noi Using the expansion (\ref{FF_H_exp}) the expression
(\ref{FLE_MSD}) can be casted as

\be
\langle
\delta^2 h\left(\vec{x},t\right)\rangle  = \frac{8k_BT(4\pi)^{-d/2}
A^{\beta}}{\gamma\Gamma(d/2)\cos\left(\pi\beta/2\right)}\int_{0}^{+\infty}d\omega\frac{1-\cos\left(\omega t\right)}{\left(\omega\right)^{1+\beta}}.
\label{FLE_MSD_1}
\ee

\noi Solving the integral we get the expression
(\ref{EW_corr_func_z>d}) obtained previously from the GEM (\ref{GEM}).

\subsection{Dynamic structure factor of fluid membranes}

In ref.~\cite{Zilman} Zilman and Granek derived the short length and
short time behavior of  the dynamic structure factor
of the fluid membranes. As mentioned in the
Introduction, the fluid
membrane dynamics  correspond to take $D=1$, $d=2$ and $z=4$ in
(\ref{GEM}), moreover the hydrodynamic friction kernel is expressed as

\be
\Lambda\left(\vec{r}\right)=\frac{1}{8\pi\xi\left|\vec{r}\right|}
\label{friction_ZG}
\ee

\noi which gives $\alpha=1$ in (\ref{hydro}), $\xi$ is the solvent
viscosity. Note that unlike the original model we set the bending
modulus $\kappa=1$. According to (\ref{hydro_FF}) and the definition
(\ref{friction_ZG})  the constant $A$ is found to be $A=1/(4\xi)$.

\noi The quantity which has been studied in ~\cite{Zilman} is the
two-point correlation function

\be
\langle \left[
  h\left(\vec{x},t\right)-h\left(\vec{x'},0\right)\right]^2\rangle=
2\left[\langle h^2\left(\vec{x},0\right)\rangle-\langle h\left(\vec{x},t\right)h\left(\vec{x'},0\right)\rangle\right]
\label{corr_ZG}
\ee

\noi which is shown to be linked to the membrane's dynamic structure
factor. The last expression can thus be recast in term of the quantity
(\ref{h_corr_FLE_2}): using the expansion (\ref{FF_H_exp}) and the numerical
values of $d$, $z$, $\alpha$ and $A$  we find

\be
\begin{array}{l}
\langle \left[
  h\left(\vec{x},t\right)-h\left(\vec{x'},0\right)\right]^2\rangle=\\
\frac{2\,
  k_BT}{3\pi(4\xi)^{2/3}}\int_{0}^{+\infty}d\omega\left\{\frac{1}{\omega^{5/3}}-\cos\left(\omega t\right)
\frac{\left(4\xi\right)^{1/3}}{8\pi\omega^{4/3}}\left|\vec{x}-\vec{x'}\right|\times\right.\\
\times \left.H_{1\,3}^{2\,1}\left[\left|\vec{x}-\vec{x'}\right|\frac{\left(4\xi\omega\right)^{1/3}}{2}\left|{\begin{array}{ccc}
&\left(\frac{2}{3},\frac{1}{6}\right)& \\
\left(\frac{-1}{2},\frac{1}{2}\right)&\left(\frac{2}{3},\frac{1}{6}\right)&\left(-\frac{1}{2},\frac{1}{2}\right)\\
\end{array} } \right.\right]\right\}
\label{corr_ZG_1}
\end{array}
\ee

\noi After changing variable
($y=\left|\vec{x}-\vec{x'}\right|(4\xi\omega)^{1/3}/2$) and making
use of the property
(\ref{1.2.5}) the following general simpler expression  is achieved
for the two-point correlation function:

\be
\begin{array}{l}
\langle \left[
  h\left(\vec{x},t\right)-h\left(\vec{x'},0\right)\right]^2\rangle=\frac{\,
  k_BT}{2\pi}\int_{0}^{+\infty}dy
\frac{\left|\vec{x}-\vec{x'}\right|^2}{y^3}\times\\
\times\left\{1-\frac{\cos\left(\left(\frac{2y}{\left|\vec{x}-\vec{x'}\right|}\right)^3 \frac{t}{4\xi}\right)}{4\pi}
H_{1\,3}^{2\,1}\left[y\left|{\begin{array}{ccc}
&\left(\frac{5}{6},\frac{1}{6}\right)& \\
\left(0,\frac{1}{2}\right)&\left(\frac{5}{6},\frac{1}{6}\right)&\left(0,\frac{1}{2}\right)\\
\end{array} } \right.\right]\right\}.
\label{corr_ZG_2}
\end{array}
\ee

\noi Although eq.(\ref{corr_ZG_2}) furnishes a compact
analitycal expression for the correlation function at any time and any
distance $\left|\vec{x}-\vec{x'}\right|$, the integral
cannot be computed esplicitely since it displays a logarithmic divergence in the
limit $y\to 0$. Again this can be seen by expanding the cosine and the Fox function
in (\ref{corr_ZG_2}) to the second order:

\be
H_{1\,3}^{2\,1}\left[y\left|{\begin{array}{ccc}
&\left(\frac{5}{6},\frac{1}{6}\right)& \\
\left(0,\frac{1}{2}\right)&\left(\frac{5}{6},\frac{1}{6}\right)&\left(0,\frac{1}{2}\right)\\
\end{array} } \right.\right] \simeq 4\pi-2\pi y^2+\pi y^4.
\label{FF_ZG_exp}
\ee

We first analyze the limit $t=0$ which corresponds to the
static correlator $\langle \left[
  h\left(\vec{x},0\right)-h\left(\vec{x'},0\right)\right]^2\rangle$,
describing the membrane's roughness.  In this case we immediately get
from (\ref{corr_ZG_2})

\be
\begin{array}{l}
\langle \left[
  h\left(\vec{x},0\right)-h\left(\vec{x'},0\right)\right]^2\rangle=\frac{\,
  k_BT}{2\pi}\int_{0}^{+\infty}dy
\frac{\left|\vec{x}-\vec{x'}\right|^2}{y^3}\times\\
\times\left\{1-\frac{1}{4\pi}
H_{1\,3}^{2\,1}\left[y\left|{\begin{array}{ccc}
&\left(\frac{5}{6},\frac{1}{6}\right)& \\
\left(0,\frac{1}{2}\right)&\left(\frac{5}{6},\frac{1}{6}\right)&\left(0,\frac{1}{2}\right)\\
\end{array} } \right.\right]\right\}.
\label{corr_ZG_3}
\end{array}
\ee

\noi On the score of the previous discussion, using the expansion
(\ref{FF_ZG_exp}) the above expression can be approximated as

\be
\begin{array}{l}
\langle \left[
  h\left(\vec{x},0\right)-h\left(\vec{x'},0\right)\right]^2\rangle\simeq\frac{\,
  k_BT}{4\pi}\left|\vec{x}-\vec{x'}\right|^2\times\\
\times\left(\int_{\left|\vec{x}-\vec{x'}\right|/L}^{1}\frac{dy}{y}-\frac{1}{2}\int_0^1dy\, y\right)
\label{corr_ZG_4}
\end{array}
\ee

\noi where $L$ stands for the long scale cut-off representing the
membrane's size. The underlying assumption in
(\ref{corr_ZG_4}) is that the major contribution to the integral in
(\ref{corr_ZG_3}) comes
from $y\leq 1$; this, in turn, justifies the lower cut-off appearing in
the first integral of (\ref{corr_ZG_4}): indeed the minimum relaxation
frequency of the membrane's bending modes will be given by
$\omega_{0}=\left(\frac{2}{L}\right)^{3}\frac{1}{4\xi}$ which
corresponds to 
$y_{0}=\left|\vec{x}-\vec{x'}\right|/L$ . Hence solving 
(\ref{corr_ZG_4}) we obtain the expression which coincides with that
found in
~\cite{Zilman} for the static correlator, i.e.

\be
\begin{array}{l}
\langle \left[
  h\left(\vec{x},0\right)-h\left(\vec{x'},0\right)\right]^2\rangle\simeq\\
\frac{\,
  k_BT}{4\pi}\left|\vec{x}-\vec{x'}\right|^2\left[\ln\left(\frac{L}{\left|\vec{x}-\vec{x'}\right|}\right)-0.25\right]
\label{corr_ZG_5}
\end{array}
\ee

\noi valid whenever $\left|\vec{x}-\vec{x'}\right|\ll L$. The value of
-0.25 of the correction term can be improved in a regular way; see
Appendix \ref{app:SS}.

\noi For the dynamic correlator  we can still consider the main
contribution to the 
integral arising from $y\leq 1$ and use (\ref{FF_ZG_exp}), then

\be
\begin{array}{l}
\langle \left[
  h\left(\vec{x},t\right)-h\left(\vec{x'},0\right)\right]^2\rangle\simeq\frac{\,
  k_BT}{2\pi}\left|\vec{x}-\vec{x'}\right|^2\times\\
\times\left[\int_{0}^{1}dy
\frac{1-\cos\left(\left(\frac{2y}{\left|\vec{x}-\vec{x'}\right|}\right)^3 \frac{t}{4\xi}\right)}{y^3}+\frac{1}{2}\int_{0}^{1}dy\frac{\cos\left(\left(\frac{2y}{\left|\vec{x}-\vec{x'}\right|}\right)^3 \frac{t}{4\xi}\right)}{y}\right].
\label{corr_ZG_6}
\end{array}
\ee

\noi Changing the variable back to $\omega$, for times in the
intermediate range
$4\xi\left(\frac{\left|\vec{x}-\vec{x'}\right|}{2}\right)^3\ll
t\ll4\xi\left(\frac{L}{2}\right)^3$ we can safely replace the upper
bound of the integrals to $\infty$ achieving

\be
\begin{array}{l}
\langle \left[
  h\left(\vec{x},t\right)-h\left(\vec{x'},0\right)\right]^2\rangle\simeq\frac{\,
  k_BT}{2\pi}\left|\vec{x}-\vec{x'}\right|^2\times\\
\times\left[\frac{(4\xi)^{-2/3}}{3}\left(\frac{4}{\left|\vec{x}-\vec{x'}\right|}\right)^2 \int_{0}^{\infty}d\omega
\frac{1-\cos\left(\omega
  t\right)}{\omega^{5/3}}+\frac{1}{6}\int_{0}^{\infty}dy\frac{\cos\left(\omega
  t\right)}{\omega}\right].
\label{corr_ZG_7}
\end{array}
\ee

\noi Thus we can compute the first integral and introduce the
infrafred cut-off $\omega_0$ to regularize the second, therefore the final
expression for the correlation function reads

\be
\begin{array}{l}
\langle \left[
  h\left(\vec{x},t\right)-h\left(\vec{x'},0\right)\right]^2\rangle\simeq\frac{\,
  k_BT}{2\pi}\left|\vec{x}-\vec{x'}\right|^2\times\\
\times\left[\frac{\Gamma(1/3)}{(4\xi)^{2/3}}\frac{t^{2/3}}{\left|\vec{x}-\vec{x'}\right|^2}-\frac{Ci\left(\frac{2t}{L^3\xi}\right)}{6}\right].
\label{corr_ZG_8}
\end{array}
\ee

\noi where $Ci$ is the cosine integral ~\cite{Abramowitz}. The result
(\ref{corr_ZG_8}) matches and amends the corresponding expression furnished in ~\cite{Zilman}.

\subsection{Donor-Acceptor correlation function in proteins}

In Refs~\cite{Kou,Min} it has been shown that the dynamics of the
donor-acceptor (D-A) distance within a protein can be mapped into the
motion of a fictitious particle obeying  a FLE with fractional
derivative of order 1/2, in presence of an harmonic potential whose 
frequency $\omega_0^2$ could be phenomenologically inferred \emph{a posteriori} from the experimental data. The  detected quantity
was the autocorrelation function of the D-A
distance $\mathbf{\Delta}_{D-A}(t)$ that was shown to display an
asymptotic Mittag-Leffler decay in accordance with the FLE
prescription. In order to recover the experimental results, in
Refs~\cite{Debnath,Dua} and ~\cite{Tang} the authors used a
respectively continuous and discrete Rouse model accounting for the
protein conformational dynamics:  this, in turn, corresponds to take
$D=3,\, d=1,\, z=2$ and
$\Lambda\left(x-x'\right)=\delta^d\left(x-x'\right)$
in (\ref{GEM}) (see Introduction). The simple Rouse model was shown to
reproduce the Mittag-Leffler decay of the $\mathbf{\Delta}_{D-A}$ autocorrelation function for large $t$ and, on the other hand, it was
shown to lead to the correct $1/2$-FLE for the D-A distance
within the framework
developed in ~\cite{Lizana,Taloni-FLE}, with the frequency $\omega_0^2\propto1/\left(x_A-x_D\right)$.  We now want to calculate the D-A
autocorrelation function without resorting to the derivation of the
FLE for the D-A distance (which will be done in Appendix
\ref{app:D-A_GLE}), but starting from the FLE (\ref{FLE}) valid for
the monomer  placed at position $x$ measured along the protein's profile.

According to  ~\cite{Debnath} we define the
$\mathbf{\Delta}_{D-A}$ autocorrelation function as

\be
\langle\mathbf{\Delta}_{D-A}(t)\cdot\mathbf{\Delta}_{D-A}(t')\rangle=3\langle\Delta_{D-A}(t)\Delta_{D-A}(t')\rangle
\label{D-A_corr}
\ee

\noi which can be rewritten in term of the spatial  positions of the donor and
acceptor sites, $h\left(x_D,t\right)$ and  $h\left(x_A,t\right)$
respectively, as

\be
\begin{array}{l}
\langle\mathbf{\Delta}_{D-A}(t)\cdot\mathbf{\Delta}_{D-A}(t')\rangle=\\
\begin{split}
6\left(\langle
h\left(x_A,t\right)h\left(x_A,t'\right)\rangle-\langle
h\left(x_A,t\right)h\left(x_D,t'\right)\rangle\right)
\end{split}
\label{D-A_corr_1}
\end{array}
\ee

\noi Therefore, putting in (\ref{h_corr_FLE_2}) the numerical value of
the parameters and changing variable ($y=\left|x_A-x_D\right|\sqrt{\left|\omega\right|}/2$) we get

\be
\begin{array}{l}
\langle\mathbf{\Delta}_{D-A}(t)\cdot\mathbf{\Delta}_{D-A}(t')\rangle=\frac{3\sqrt{2}\,
  k_BT}{\pi}\left|x_A-x_D\right|\times\\
\times\int_{0}^{+\infty}dy\frac{\cos\left(\left(\frac{2y}{\left|x_A-x_D\right|}\right)^2 \left|t-t'\right|\right)}
{y^2}\times\\
\times\left\{1-\frac{1}{2^{3/2}\sqrt{\pi}}
H_{1\,3}^{2\,1}\left[y\left|{\begin{array}{ccc}
&\left(\frac{3}{4},\frac{1}{4}\right)& \\
\left(0,\frac{1}{2}\right)&\left(\frac{3}{4},\frac{1}{4}\right)&\left(\frac{1}{2},\frac{1}{2}\right)\\
\end{array} } \right.\right]\right\}.
\label{D-A_corr_2}
\end{array}
\ee

\noi Expanding the $H$-function in the former expression gives

\be
H_{1\,3}^{2\,1}\left[y\left|{\begin{array}{ccc}
&\left(\frac{3}{4},\frac{1}{4}\right)& \\
\left(0,\frac{1}{2}\right)&\left(\frac{3}{4},\frac{1}{4}\right)&\left(\frac{1}{2},\frac{1}{2}\right)\\
\end{array} } \right.\right] \simeq 2^{3/2}\sqrt{\pi}\left[1-2y^{2}\right].
\label{D-A_corr_exp}
\ee

\noi which guarantees that the integral in (\ref{D-A_corr_2}) does not present any divergence: we then seek for its explicit solution. First we express the Fox function's Mellin transform following the definition (\ref{FF_def}): according to  (\ref{Mellin_def}) it is found to be

\be
\chi(s)=\frac{2^{3/2}\sqrt{\pi}\Gamma(s)\left[\sin\left(\frac{\pi s}{4}\right)+\cos\left(\frac{\pi s}{4}\right)\right]}{2^s}
\label{D-A_corr_Mellin}
\ee

\noi which can be inverted ~\cite{Obertinger} to give

\be
\begin{array}{l}
H_{1\,3}^{2\,1}\left[y\left|{\begin{array}{ccc}
&\left(\frac{3}{4},\frac{1}{4}\right)& \\
\left(0,\frac{1}{2}\right)&\left(\frac{3}{4},\frac{1}{4}\right)&\left(\frac{1}{2},\frac{1}{2}\right)\\
\end{array} } \right.\right]=\\
2^{3/2}\sqrt{\pi}e^{-\sqrt{2}y}\left(\sin\sqrt{2}y +\cos\sqrt{2}y\right).
\label{D-A_corr_FF_expr}
\end{array}
\ee

\noi Plugging (\ref{D-A_corr_FF_expr}) in (\ref{D-A_corr_2}) and integrating  by parts, we obtain

\be
\begin{array}{l}
\langle\mathbf{\Delta}_{D-A}(t)\cdot\mathbf{\Delta}_{D-A}(t')\rangle=\frac{3\sqrt{2}\,
  k_BT}{\pi}\left|x_A-x_D\right|\times\\
\times \left[  +2C\int_0^{\infty}dy \sin \left(Cy^2\right) e^{-\sqrt{2}y}\left(\sin\sqrt{2}y +\cos\sqrt{2}y\right)+\right.\\
+2\sqrt{2}\int_0^{\infty}dy \frac{\sin\sqrt{2}y\cos \left(Cy^2\right)}{y}e^{-\sqrt{2}y}-2C\int_0^{\infty}dy \sin Cy^2\left.\right]
\label{D-A_corr_3}
\end{array}
\ee

\noi where $C=\frac{4\left|t-t'\right|}{\left|x_A-x_D\right|^2}$. Let's analyze the three integrals within the square brackets. First we consider the following general expression

\be
\int_0^{\infty}dy  y e^{-\beta y} \sin\left(Cy^2\right)\sin\left(\beta y\right)=\sqrt{\frac{\pi}{2C^3}}\,\frac{\beta}{4}e^{-\frac{\beta^2}{2C}}.
\label{D-A_I_1}
\ee

\noi We then integrate with respect to $\beta$ both the RHS and the LHS of (\ref{D-A_I_1}) to achieve ~\cite{Gradshtein}

\be
\int_0^{\infty}dy   e^{-\beta y} \sin\left(Cy^2\right)\left(\sin\left(\beta y\right)+\cos\left(\beta y\right)\right)=\sqrt{\frac{\pi}{C}}\frac{e^{-\frac{\beta^2}{2C}}}{2^{3/2}}
\label{D-A_I_1_2}
\ee

\noi Multiplying both sides by $2C$ and setting $\beta=\sqrt{2}$ eq.(\ref{D-A_I_1_2}) yields

\be
\begin{array}{l}
2C\int_0^{\infty}dy   \sin\left(Cy^2\right)e^{-\sqrt{2} y}\left(\sin\left(\sqrt{2} y\right)+\cos\left(\sqrt{2} y\right)\right)=\\
\sqrt{\frac{\pi C}{2}}e^{-\frac{1}{C}}.
\label{D-A_I_1_3}
\end{array}
\ee

\noi We next consider the second integral in (\ref{D-A_corr_3}), hereby named $I_2(C)$, and differentiate it with respect to C, i.e.

\be
\frac{d}{dC}I_2(C)=-\sqrt{\frac{\pi}{C^3}}\frac{e^{-\frac{1}{C}}}{\sqrt{2}}.
\label{D-A_I_2_1}
\ee

\noi The former differential equation can be solved thanks to the initial condition ~\cite{Gradshtein}

\be
I_2(0)=2\sqrt{2}\int_0^{\infty}dy \frac{\sin\sqrt{2}y}{y}e^{-\sqrt{2}y}=\frac{\pi}{\sqrt{2}}
\label{D-A_I_2_2}
\ee

\noi yielding

\be
I_2(C)=\frac{\pi}{\sqrt{2}}\,erf\left[\frac{1}{\sqrt{C}}\right]
\label{D-A_I_2_3}
\ee

\noi where erf denotes the error function ~\cite{Abramowitz}.

\noi The third integral in (\ref{D-A_corr_3}) can be evaluated effortlessly ~\cite{Gradshtein}

\be
-2C\int_0^{\infty}dy \sin Cy^2=-\sqrt{\frac{\pi C}{2}}
\label{D-A_I_3_1}.
\ee

\noi Putting (\ref{D-A_I_1_3}), (\ref{D-A_I_2_3}) and (\ref{D-A_I_3_1}) in (\ref{D-A_corr_3}) and substituting the expression of $C$, the final form of the D-A autocorrelation function is achieved:

\be
\begin{array}{l}
\langle\mathbf{\Delta}_{D-A}(t)\cdot\mathbf{\Delta}_{D-A}(t')\rangle=\\
\frac{3\sqrt{2}\,
  k_BT}{\pi}\left|x_A-x_D\right|
\times\left\{\sqrt{\frac{2\pi\left|t-t'\right|}{\left|x_A-x_D\right|^2}}\left(e^{-\frac{\left|x_A-x_D\right|^2}{4\left|t-t'\right|}}-1\right)+\right.\\
\left.+\frac{\pi}{\sqrt{2}}\,erf\left[\frac{\left|x_A-x_D\right|}{2}\sqrt{\frac{1}{\left|t-t'\right|}}\right]\right\}
\label{D-A_corr_4}
\end{array}
\ee

\noi which is exactly the expression found in ref~\cite{Lizana}. Moreover (\ref{D-A_corr_4}) recovers the asymptotic decay of the autocorrelation function $C_x(t)$ observed in the experiments ~\cite{Kou,Min} that indeed was found to be

\be
C_x(t)= \frac{k_BT}{\omega_0^2} E_{1/2,1}\left[-\sqrt{\frac{t}{t_0}}\right]\sim \frac{k_BT}{\omega_0^2 \sqrt{\pi}}\sqrt{\frac{t_0}{t}},
\label{Corr_exp}
\ee

\noi where $\omega_0$ is the charachteristic frequency of the potential and $t_0=\left(\frac{\xi}{\omega_0^2}\right)^2$, with $\xi$ generalized damping coefficient.  On the other hand the expression (\ref{D-A_corr_4}) asymptotically attains the form

\be
\langle\Delta_{D-A}(0)\Delta_{D-A}(t)\rangle \sim \frac{3k_BT\left(x_A-x_D\right)^2}{\sqrt{\pi}}\frac{1}{\sqrt{t}}
\label{Corr_our}
\ee

\noi which matches the former expression once $t_0=\left|x_A-x_D\right|^2$ and $\omega_0\propto 1/\sqrt{\left|x_A-x_D\right|}$ (see Appendix \ref{app:D-A_GLE}). We note that, whereas  (\ref{D-A_corr_4}) does not furnish a \emph{pure} Mittag-Leffler decay, the comparison with the real experimental data is very good even at short times, as shown in ~\cite{Tang}. Moreover, the expression (\ref{D-A_corr_4}) and the corresponding single-file's ~\cite{Lizana} provide a compact and elegant representation of the D-A distance correlation function, reproducing the results obtained formerly starting from the same Rouse's protein models ~\cite{Debnath,Tang}.

\subsection{Correlation function in fluctuating interfaces}

In fluctuating interfaces ~\cite{interfaces,Krug,surfaces}, often referred as rough surfaces, the height $h(\vec{x},t)$ obeys to the following generalized growth equation (generalized elastic line)

\begin{equation}
\frac{\partial}{\partial t}h\left(\vec{x},t\right)=\frac{\partial^z }{\partial\left|\vec{x}\right|^z }h(\vec{x},t)+\eta\left(\vec{x},t\right).
\label{GEL}
\end{equation}

\noi Accordingly $\Lambda\left(\vec{x}-\vec{x'}\right)=\delta^d\left(\vec{x}-\vec{x'}\right)$ while $z$ and $d$ are left unspecified. The growth of the interface is charachterized by two different growth exponents: the dynamical exponent $z$
and the roughness exponent $\zeta_r$. These two exponents control the correlations among the neighboring heights: the correlation between two different sites $\vec{x}$ and $\vec{x'}$ grows according to $\left|\vec{x}-\vec{x'}\right|\propto t^{1/z}$, while the difference between  the corresponding heights behaves like $\left|h\left(\vec{x},t\right)-h\left(\vec{x'},t\right)\right|\propto \left|\vec{x}-\vec{x'}\right|^{\zeta_r}$. A scaling argument ~\cite{Barabasi} requires that 

\be
\zeta_r=\frac{z-d}{2}
\label{zeta_fluct}
\ee

\noi since eq.(\ref{GEL}) must be invariant under scale transformation. However we can obtain both the scaling form and the exponents in a rigourous way, by studying the following correlation function

\be
\begin{array}{l}
\langle \left[
  h\left(\vec{x},t\right)-h\left(\vec{x},0\right)\right]\left[
  h\left(\vec{x'},t'\right)-h\left(\vec{x'},0\right)\right]\rangle=\\
\langle h\left(\vec{x},t\right)h\left(\vec{x'},t'\right)\rangle-\langle h\left(\vec{x},t\right)h\left(\vec{x'},0\right)\rangle-\\
-\langle h\left(\vec{x'},t'\right)h\left(\vec{x},0\right)\rangle+\langle h\left(\vec{x},0\right)h\left(\vec{x'},0\right)\rangle.
\label{corr_BS}
\end{array}
\ee

\noi Thanks to (\ref{h_corr_FLE_2}) the former function reads (here $\alpha=d$, $\gamma=2z$ and $\beta=(z-d)/z$)

\be
\begin{array}{l}
\langle \left[
  h\left(\vec{x},t\right)-h\left(\vec{x},0\right)\right]\left[ h\left(\vec{x'},t'\right)-h\left(\vec{x'},0\right)\right]\rangle=\\
\frac{k_BT}{4^d\pi^{d/2+1}z}\left|\vec{x}-\vec{x'}\right|^{d}
\int_{0}^{+\infty}d\omega\, \omega^{2\frac{d-z}{z}}\times\\
\times H_{1\,3}^{2\,1}\left[\frac{\left|\vec{x}-\vec{x'}\right|\omega^{\frac{1}{z}}}{2}\left|{\begin{array}{ccc}
&\left(\frac{z-d}{z},\frac{1}{2z}\right)& \\
\left(-\frac{d}{2},\frac{1}{2}\right)&\left(\frac{z-d}{d},\frac{1}{2z}\right)&\left(1-d,\frac{1}{2}\right)\\
\end{array} } \right.\right]\times\\
\times \left[\cos\left(\omega\left|t-t'\right|\right)-\cos\left(\omega t\right)-\cos\left(\omega t'\right)+1
\right].
\label{corr_BS_1}
\end{array}
\ee

\noi Changing variable $\left(y=\left|\vec{x}-\vec{x'}\right|\omega^{\frac{1}{z}}/2\right)$
and using (\ref{1.2.5}) the expression (\ref{corr_BS_1})
can be recasted as

\be
\begin{array}{l}
\langle \left[
  h\left(\vec{x},t\right)-h\left(\vec{x},0\right)\right]\left[ h\left(\vec{x'},t'\right)-h\left(\vec{x'},0\right)\right]\rangle=\\
\frac{k_BT}{2^z\pi^{d/2+1}}\left|\vec{x}-\vec{x'}\right|^{z-d}
\int_{0}^{+\infty}dy\times\\
\times H_{1\,3}^{2\,1}\left[y\left|{\begin{array}{ccc}
&\left(\frac{z-1}{2z},\frac{1}{2z}\right)& \\
\left(\frac{d-z-1}{2},\frac{1}{2}\right)&\left(\frac{z-1}{2z},\frac{1}{2z}\right)&\left(\frac{1-z}{2},\frac{1}{2}\right)\\
\end{array} } \right.\right]\times\\
\times \left[\cos\left((2y)^z\frac{\left|t-t'\right|}{\left|\vec{x}-\vec{x'}\right|^{z}}\right)-\cos\left((2y)^z\frac{t}{\left|\vec{x}-\vec{x'}\right|^{z}}\right)-\right.\\
\left.-\cos\left((2y)^z\frac{t'}{\left|\vec{x}-\vec{x'}\right|^{z}}\right)+1
\right].
\label{corr_BS_2}
\end{array}
\ee

\noi Eq.(\ref{corr_BS_2}) is the generalization of a well-known scaling formula obtained for the two-point two-time correlation function of the Edward-Wilkinson chain ~\cite{Barabasi,Natterman}. On the other hand the width of a growing surface obeys the \emph{Family-Vicsek} scaling relation ~\cite{Family}:

\be
w(L,t)\equiv \sqrt{\frac{1}{L^d}\int_{L^d}d\vec{x}\,\left[h\left(\vec{x},t\right)-\langle h\left(\vec{x},t\right)\rangle\right]^2}\sim L^{\zeta_r}g\left(\frac{t}{L^z}\right)
\label{Family_Vicsek}
\ee

\noi where $L$ is the maximum size of the system ($L^d$ is the total volume) and $g$ is a scaling function. Comparing (\ref{Family_Vicsek}) with (\ref{corr_BS_2}) we find  both the scaling expression and the correct value of the scaling exponents, i.e Eq.(\ref{zeta_fluct}). Moreover we want to stress that, to the knowledge of the authors, this is the first time that a scaling function appearing in the surface growth is given in an explicit form. 

\noi The integrals entering Eqs(\ref{corr_BS}) and (\ref{corr_BS_1}) can be performed with the use of the $H$-functions' properties, and the result is expressed in terms of their combination. Here we analyze only the simple case of the  Edward-Wilkinson chain, thus we set $z=2$ and $d=1$ in (\ref{corr_BS_2}) ~\cite{footnote}. Thanks to (\ref{D-A_corr_FF_expr}) the correlation function gets the expression

\be
\begin{array}{l}
\langle \left[
  h\left(x,t\right)-h\left(x,0\right)\right]\left[ h\left(x',t'\right)-h\left(x',0\right)\right]\rangle=\\
\frac{k_BT}{\sqrt{2}\pi}\left|x-x'\right|
\int_{0}^{+\infty}dy\frac{e^{-\sqrt{2}y}}{y^2}\left[\sin(\sqrt{2}y)+\cos(\sqrt{2}y)\right]\times \\
\times\left[\cos\left(Ay^2\right)-\cos\left(By^2\right)-\cos\left(Cy^2\right)+1
\right].
\label{corr_BS_3}
\end{array}
\ee

\noi where $A=4\frac{\left|t-t'\right|}{\left|x-x'\right|^{2}}$, $B=4\frac{t}{\left|x-x'\right|^{2}}$ and $C=4\frac{t'}{\left|x-x'\right|^{2}}$. Making use of the integrals (\ref{D-A_I_1_3}) and (\ref{D-A_I_2_3}) we achieve the final scaling form

\be
\begin{array}{l}
\langle \left[
  h\left(x,t\right)-h\left(x,0\right)\right]\left[ h\left(x',t'\right)-h\left(x',0\right)\right]\rangle=\\
\frac{k_BT}{\sqrt{2}}\left|x-x'\right|\left[f\left(\frac{t}{\left|x-x'\right|^{2}}\right)+f\left(\frac{t'}{\left|x-x'\right|^{2}}\right)-f\left(\frac{\left|t-t'\right|}{\left|x-x'\right|^{2}}\right)\right]\\
f(u)=\sqrt{\frac{2u}{\pi}}e^{-\frac{1}{4u}}-\frac{1}{\sqrt{2}}\,erfc\left(\sqrt{\frac{1}{4u}}\right).
\label{corr_BS_4}
\end{array}
\ee

\noi It is straightforward to verify that

\be
f(u)\left\{
\begin{array}{ccc}
\sim 8u^{3/2}e^{-\frac{1}{4u}} & & u\to 0\\
\sim \sqrt{u} & & u\to\infty.
\end{array}
\label{corr_BS_scaling}
\right.
\ee

\section{Summary and concluding remarks}

In this article we have shown how the Markovian representation of the system's dynamics furnished by the GEM (\ref{GEM}) is equivalent to the non-Markovian description of the tracer's dynamics given in terms of the FLE (\ref{FLE}). Firstly we want stress that the introduced FLE describes the time evolution of the randomm field, which depends not only upon time (as the usual FLE do) but also on space variable. Indeed, although the FLE reproduces the anomalous stochastic motion of the field $h(\vec{x},t)$ at a given position $\vec{x}$ disregarding the remaining systems' dynamics, the internal spatial correlations appear through the noise term which is correlated in time \emph{and} in space. This is the novelty of our approach: strictly speaking we don't
loose any information in passing from the Markovian GEM (\ref{GEM}) to
the non-Markovian FLE (\ref{FLE}), on the contrary we rather ease the
calculations! Indeed, the appearance of the Fox function in the two-point two-time correlation function, as well as in the noise correlation function, makes the computation of any physical observable relatively easy, if resorting to the few general Fox function's properties. Moreover, as in the case of fluctuating interfaces, the proposed framework allows the expression of  the statistical quantities in an explicit analytical and elegant form involving the $H$-functions.

\subsection*{Acknowledgments}
ACh acknowledges financial support from European Commission via MC
IIF grant No.219966 LeFrac.

%%%%%%%%%%%%%%%%%%%%%%%%%%%%%%%%%%%%%%%%%%%%%%%%%%%%%%%%%%%%%%%%%%%%%%%%%%%%%%%
%
%                      A P P E N D I X A
%
%%%%%%%%%%%%%%%%%%%%%%%%%%%%%%%%%%%%%%%%%%%%%%%%%%%%%%%%%%%%%%%%%%%%%%%%%%%%%%%
\appendix
\section{Hydrodynamic term's Fourier transform: case $\alpha=d$}
\label{app:hydro}

If the hydrodynamic term is expressed as

\be
\Lambda\left(\vec{r}\right)=\frac{1}{a^d+\left|\vec{r}\right|^d}
\label{app:hydro:general}
\ee

\noi its Fourier transform expression reads

\be
\Lambda\left(\vec{q}\right)=(2\pi)^{d/2}\left|\vec{q}\right|^{1-d/2}\int_0^{\infty}dr\frac{r^{d/2}}{a^{d}+r^{d}}J_{d/2-1}\left(\left|\vec{q}\right|r\right),
\label{app:hydro:general_FT}
\ee

\noi where $r$ stands for $\left|\vec{r}\right|$ and $J_{d/2-1}$ represents the Bessel
function of order $d/2-1$. In what follows
we will provide the exact result for the 1, 2, and 3-dimensional
cases.

\emph{i)} $d=1$. In this case the expression (\ref{app:hydro:general_FT})
takes the simple form

\be
\Lambda\left(q\right)=2\int_0^{\infty}dr\frac{\cos(qr)}{a+r}.
\label{app:hydro:general_FT_d=1_1}
\ee

\noi Solving Eq.(\ref{app:hydro:general_FT_d=1_1}) we get the following result

\be
\Lambda\left(q\right)=2\left[\sin(qa)\left(\frac{\pi}{2}-Si(qa)\right)-\cos(qa)Ci(qa)\right],
\label{app:hydro:general_FT_d=1_2}
\ee

\noi where $Si$ and $Ci$ represent respectively the sine and cosine
integrals  ~\cite{Abramowitz} .

\emph{ii)} $d=2$. We have

\be
\Lambda\left(q\right)=2\pi\int_0^{\infty}dr\frac{r}{a^2+r^2}J_{0}\left(\left|\vec{q}\right|r\right)
\label{app:hydro:general_FT_d=2_1}
\ee

\noi which gives ~\cite{Gradshtein}

\be
\Lambda\left(\vec{q}\right)=2\pi K_0\left(\left|\vec{q}\right|a\right),
\label{app:hydro:general_FT_d=2_2}
\ee

\noi where $K_0$ stands for the modified Bessel function
of 0-th order ~\cite{Abramowitz}.

\emph{iii)} $d=3$. Eq.(\ref{app:hydro:general_FT}) takes the following
form,

\be
\Lambda\left(\vec{q}\right)=\frac{(2\pi)^{3/2}}{\sqrt{\left|\vec{q}\right|}}\int_0^{\infty}dr\frac{r^{3/2}}{a^3+r^3}J_{1/2}\left(\left|\vec{q}\right|r\right),
\label{app:hydro:general_FT_d=3_1}
\ee

\noi which can be rewritten as

\be
\Lambda\left(\vec{q}\right)=\frac{4\pi}{\left|\vec{q}\right|}\int_0^{\infty}dr\frac{r}{a^3+r^3}\sin\left(\left|\vec{q}\right|r\right).
\label{app:hydro:general_FT_d=3_2}
\ee

\noi The integral in the previous expression can be splitted into the
sum of two
contributions,
i.e. $\Lambda\left(\vec{q}\right)=\frac{4\pi}{\left|\vec{q}\right|}\left(I_1\left(\vec{q}\right)-I_2\left(\vec{q}\right)\right)$ with

\be
\left\{
\begin{array}{lll}
I_1\left(\vec{q}\right) & = &\int_0^{\infty}dr\frac{\sin\left(\left|\vec{q}\right|r\right)}{(r-x)(r-x*)}\\
I_2\left(\vec{q}\right) & = &a\int_0^{\infty}dr\frac{\sin\left(\left|\vec{q}\right|r\right)}{\left(r+a\right)(r-x)\left(r-x^*\right)},
\label{app:hydro:general_FT_d=3_3}
\end{array}
\right.
\ee

\noi where $x=ae^{i\pi/3}$ and $x^*=ae^{-i\pi/3}$. We
first study $I_1\left(\vec{q}\right)$, which can be easily transformed in

\be
I_1\left(\vec{q}\right)=\frac{1}{ia\sqrt{3}}\left[\int_{-x}^{\infty}dy\frac{\sin\left(\left|\vec{q}\right|(y+x)\right)}{y}-c.c.\right]
\label{app:hydro:general_FT_d=3_4}
\ee

\noi where for $c.c$ we denoted the complex conjugated of the first
integral in the square brackets. After a bit
of algebra we achieve for the final form of $I_1\left(\vec{q}\right)$:

\be
\left\{
\begin{array}{lll}
I_1\left(\vec{q}\right) &  = & \frac{1}{ia\sqrt{3}}\left[f(x)-f\left(x^*\right)\right]\\
f(x)  & = &
\cos\left(\left|\vec{q}\right|x\right)\left[Si\left(\left|\vec{q}\right|x\right)+\frac{\pi}{2}\right]-\sin\left(\left|\vec{q}\right|x\right)Ci\left(\left|\vec{q}\right|x\right)
\label{app:hydro:general_FT_d=3_5}
\end{array}
\right.
\ee

\noi The integral $I_2\left(\vec{q}\right)$ can be manipulated to get

\be
\begin{array}{lll}
I_2\left(\vec{q}\right) & = & \frac{1}{i\sqrt{3}}\left\{\frac{1}{\left(a+x\right)}\left[\int_{-x}^{\infty}dy\frac{\sin\left(\left|\vec{q}\right|(y+x)\right)}{y}-\right.\right.\\
 & & \left.\left.-\int_{a}^{\infty}dy\frac{\sin\left(\left|\vec{q}\right|(y-a\right)}{y}\right]-c.c.\right\},
\end{array}
\label{app:hydro:general_FT_d=3_6}
\ee

\noi where $c.c.$  this time represents the complex conjugate of the whole
expression in the curly brackets. According to the same procedure used for
the former integral
$I_1\left(\vec{q}\right)$ we achieve

\be
\left\{
\begin{array}{lll}
I_2\left(\vec{q}\right) &  = & \frac{g\left(a,x\right)}{i\sqrt{3}\left(a+x\right)}-\frac{g\left(a,x^*\right)}{i\sqrt{3}\left(a+x^*\right)}\\
g\left(a,x\right)  & = & f(x)+
\cos\left(\left|\vec{q}\right|a\right)\left[Si\left(\left|\vec{q}\right|a\right)-\frac{\pi}{2}\right]-\\
& & -\sin\left(\left|\vec{q}\right|a\right)Ci\left(\left|\vec{q}\right|a\right)
\label{app:hydro:general_FT_d=3_7}.
\end{array}
\right.
\ee

We recover immediately the asymptotic expression
$\Lambda\left(\vec{q}\right)\sim
\frac{2\pi^{d/2}}{\Gamma\left(d/2\right)}\ln\left(\frac{1}{\left|\vec{q}\right|a}\right)$
by expanding
Eqs(\ref{app:hydro:general_FT_d=1_2}),(\ref{app:hydro:general_FT_d=2_2})
and the solution of (\ref{app:hydro:general_FT_d=3_2}) for small $\left|\vec{q}\right|$.

%%%%%%%%%%%%%%%%%%%%%%%%%%%%%%%%%%%%%%%%%%%%%%%%%%%%%%%%%%%%%%%%%%%%%%%%%%%%%%%
%
%                      A P P E N D I X B
%
%%%%%%%%%%%%%%%%%%%%%%%%%%%%%%%%%%%%%%%%%%%%%%%%%%%%%%%%%%%%%%%%%%%%%%%%%%%%%%%
\section{Fox function appearance}
\label{app:FF}

The Fox functions are defined as ~\cite{Fox,Mathai,Hilfer,Prudnikov}

\be
H_{p\,q}^{m\,n}\left[y\left|{\begin{array}{ccc}
(a_1,A_1)&...&(a_p,A_p)\\
(b_1,B_1)&...&(b_q,B_q)\\
\end{array} } \right.\right]=\frac{1}{2\pi i}\int_L\chi(s)y^{-s}ds
\label{FF_def}
\ee

\noi with $1\leq m\leq q$, $0\leq n\leq p$.  $\chi(s)$ represents
the Mellin transform which takes the following form

\be
\chi(s)=\frac{\prod_{j=1}^{m}\Gamma\left(b_j+B_js\right)\prod_{j=1}^{n}\Gamma\left(1-a_j-A_js\right)}{\prod_{j=m+1}^{q}\Gamma\left(1-b_j-B_js\right)\prod_{j=n+1}^{p}\Gamma\left(a_j+A_js\right)}.
\label{Mellin_def}
\ee

\noi where $A_j$ and $B_j$ are positive numbers while  $a_j$ and
$b_j$ are complex. Empty products are interpreted as being unity.

In the expression (\ref{phi_2}) the function appearing in the denominator can be
expressed as a Fox function, see eq.(\ref{FF_appearence}). Indeed it is sufficient to notice that the
Mellin transform of $\frac{1}{1+y^{\gamma}}$ is given by

\be
\chi(s)=\int_0^{\infty}\frac{1}{1+y^{\gamma}}y^{s-1}ds=\frac{\Gamma\left(s/\gamma\right)\Gamma\left(1-s/\gamma\right)}{\gamma}
\label{Mellin_phi_2}
\ee

\noi which matches the definition (\ref{Mellin_def}) if and only if
$m=n=p=q=1$, $a_1=b_1=0$ and $A_1=B_1=1/\gamma$.

%%%%%%%%%%%%%%%%%%%%%%%%%%%%%%%%%%%%%%%%%%%%%%%%%%%%%%%%%%%%%%%%%%%%%%%%%%%%%%%
%
%                      A P P E N D I X C
%
%%%%%%%%%%%%%%%%%%%%%%%%%%%%%%%%%%%%%%%%%%%%%%%%%%%%%%%%%%%%%%%%%%%%%%%%%%%%%%%
\section{Fox function properties}
\label{app:FF_prop}

In this Section we enumerate the properties of the Fox function
(\ref{FF_def}) that we use throughout our analysis. This list is  not an exhaustive compendium of the Fox functions properties, for which the reader could
refer to ~\cite{Mathai,Hilfer,Prudnikov}.

\noi For convenience in this Section we adopt the following short notation

\be
H_{p\,q}^{m\,n}\left[y\left|{\begin{array}{ccc}
(a_1,A_1)&...&(a_p,A_p)\\
(b_1,B_1)&...&(b_q,B_q)\\
\end{array} } \right.\right]=H_{p\,q}^{m\,n}\left[y\left|{\begin{array}{c}
\left[a_p,A_p\right]\\
\left[b_q,B_q\right]\\
\end{array} } \right.\right].
\label{FF_short}
\ee

The useful rules are hereafter listed:

\be
H_{p\,q}^{m\,n}\left[y\left|{\begin{array}{c}
\left[a_p,A_p\right]\\
\left[b_q,B_q\right]\\
\end{array} } \right.\right]=
H_{q\,p}^{n\,m}\left[\frac{1}{y}\left|{\begin{array}{c}
\left[1-b_q,B_q\right]\\
\left[1-a_p,A_p\right]\\
\end{array} } \right.\right]
\label{1.2.3}
\ee

\be
y^{\sigma}H_{p\,q}^{m\,n}\left[y\left|{\begin{array}{c}
\left[a_p,A_p\right]\\
\left[b_q,B_q\right]\\
\end{array} } \right.\right]=
H_{p\,q}^{m\,n}\left[y\left|{\begin{array}{c}
\left[a_p+\sigma A_p,A_p\right]\\
\left[b_q+\sigma B_q,B_q\right]\\
\end{array} } \right.\right]
\label{1.2.5}
\ee

\be
\begin{array}{c}
\int_0^{\infty}dy\,y^{\alpha-1}J_{\nu}(\sigma y)H_{p\,q}^{m\,n}\left[\omega y^r\left|{\begin{array}{c}
\left[a_p,A_p\right]\\
\left[b_q,B_q\right]\\
\end{array} } \right.\right]=\\
\frac{2^{\alpha-1}}{\sigma^{\alpha}}H_{p+2\,,q}^{m\,,n+1}\left[\omega \left(\frac{2}{\sigma}\right)^r\left|{\begin{array}{c}
\left(1-\frac{\alpha+\nu}{2},\frac{r}{2}\right)\left[a_p,A_p\right]\left(1-\frac{\alpha-\nu}{2},\frac{r}{2}\right)\\
\left[b_q,B_q\right]\\
\end{array} } \right.\right]
\end{array}
\label{pr.5}
\ee

\be
\begin{array}{c}
\int_0^{\infty}dy\,y^{\alpha-1}\cos(\sigma y)H_{p\,q}^{m\,n}\left[\omega y^r\left|{\begin{array}{c}
\left[a_p,A_p\right]\\
\left[b_q,B_q\right]\\
\end{array} } \right.\right]=\\
\frac{2^{\alpha-1}\sqrt{\pi}}{\sigma^{\alpha}}H_{p+2\,,q}^{m\,,n+1}\left[\omega \left(\frac{2}{\sigma}\right)^r\left|{\begin{array}{c}
\left(\frac{2-\alpha}{2},\frac{r}{2}\right)\left[a_p,A_p\right]\left(\frac{1-\alpha}{2},\frac{r}{2}\right)\\
\left[b_q,B_q\right]\\
\end{array} } \right.\right]
\end{array}
\label{pr.4}
\ee

\be
\begin{array}{c}
y^{r}\frac{d^r}{dy^r}H_{p\,q}^{m\,n}\left[ y^{\delta}\left|{\begin{array}{c}
\left[a_p,A_p\right]\\
\left[b_q,B_q\right]\\
\end{array} } \right.\right]=
H_{p+1\,,q+1}^{m\,,n+1}\left[y^{\delta}\left|{\begin{array}{c}
\left(0,\delta\right)\left[a_p,A_p\right]\\
\left[b_q,B_q\right]\left(r,\delta\right)\\
\end{array} } \right.\right]
\end{array}
\label{pr.diff}
\ee

\be
\begin{array}{l}
H_{p\,q}^{m\,n}\left[y\left|{\begin{array}{c}
\left[a_p,A_p\right]\\
\left[b_q,B_q\right]\\
\end{array} }
  \right.\right]=\sum_{i=1}^{m}\sum_{k=0}^{\infty}c_{ik}\frac{(-1)^{k}}{k!B_i}y^{\frac{b_i+k}{B_i}}\\
c_{ik}=\frac{\prod_{j=1,j\neq i}^{m}\Gamma\left(b_j-\frac{(b_i+k)B_j}{B_i}\right)\prod_{j=1}^{n}\Gamma\left(1-a_j+\frac{(b_i+k)A_j}{B_i}\right)}{\prod_{j=m+1}^{q}\Gamma\left(1-b_j+\frac{(b_i+k)B_j}{B_i}\right)\prod_{j=n+1}^{p}\Gamma\left(a_j-\frac{(b_i+k)A_j}{B_i}\right)}
\label{exp}
\end{array}
\ee 

\noi This expansion is valid whenever
$\sum_{i=1}^qB_i-\sum_{i=1}^pA_i\geq 0$ or $\sum_{i=1}^qB_i-\sum_{i=1}^pA_i< 0$ and $\sum_{i=1}^nA_i-\sum_{i=n+1}^pA_i+sum_{i=1}^mB_i-sum_{i=m+1}^qB_i> 0$. Empty products are interpreted as being unity.

%%%%%%%%%%%%%%%%%%%%%%%%%%%%%%%%%%%%%%%%%%%%%%%%%%%%%%%%%%%%%%%%%%%%%%%%%%%%%%%
%
%                      A P P E N D I X D
%
%%%%%%%%%%%%%%%%%%%%%%%%%%%%%%%%%%%%%%%%%%%%%%%%%%%%%%%%%%%%%%%%%%%%%%%%%%%%%%%
\section{Static correlator for fluid membranes }
\label{app:SS}

In this appendix we show how to get a general expression for the static
correlator (\ref{corr_ZG_3}). Indeed using the property
(\ref{pr.diff}), eq.(\ref{corr_ZG_3}) becomes

\be
\begin{array}{l}
\langle \left[
  h\left(\vec{x},0\right)-h\left(\vec{x'},0\right)\right]^2\rangle=\frac{\,
  k_BT}{2\pi}\left|\vec{x}-\vec{x'}\right|^2\left\{0.25-\right.\\
\left.
-\frac{1}{8\pi}\int_{0}^{+\infty}\frac{dy}{y^3}H_{2\,4}^{2\,2}\left[y\left|{\begin{array}{cccc}
\left(0,1\right)&\left(\frac{5}{6},\frac{1}{6}\right)& \\
\left(0,\frac{1}{2}\right)&\left(\frac{5}{6},\frac{1}{6}\right)&\left(0,\frac{1}{2}\right)&\left(1,1\right)\\
\end{array} } \right.\right]\right\}.
\label{app:corr_ZG_1}
\end{array}
\ee

\noi Introducing the lower cut-off
$y_0=\left|\vec{x}-\vec{x'}\right|/L$ and extracting the logarithmic
term we can recast the previous expression as

\be
\begin{array}{l}
\langle \left[
  h\left(\vec{x},0\right)-h\left(\vec{x'},0\right)\right]^2\rangle=\\
\frac{k_BT}{4\pi}\left|\vec{x}-\vec{x'}\right|^2\left\{\ln\left(\frac{L}{\left|\vec{x}-\vec{x'}\right|}\right)
+0.5-\frac{1}{4\pi}\int_{\left|\vec{x}-\vec{x'}\right|/L}^{1}\frac{dy}{y^3}\times\right.\\
\left.
\times\left(4\pi
  y^2+H_{2\,4}^{2\,2}\left[y\left|{\begin{array}{cccc}
\left(0,1\right)&\left(\frac{5}{6},\frac{1}{6}\right)& \\
\left(0,\frac{1}{2}\right)&\left(\frac{5}{6},\frac{1}{6}\right)&\left(0,\frac{1}{2}\right)&\left(1,1\right)\\
\end{array} } \right.\right.\right]\right)-\\
\left.-\frac{1}{4\pi}\int_{1}^{\infty}\frac{dy}{y^3}H_{2\,2}^{2\,4}\left[y\left|{\begin{array}{cccc}
\left(0,1\right)&\left(\frac{5}{6},\frac{1}{6}\right)& \\
\left(0,\frac{1}{2}\right)&\left(\frac{5}{6},\frac{1}{6}\right)&\left(0,\frac{1}{2}\right)&\left(1,1\right)\\
\end{array} } \right.\right]\right\}.
\label{app:corr_ZG_2}
\end{array}
\ee

\noi The last term in the brackets is transformed with the use of (\ref{1.2.3})
and (\ref{1.2.5}), getting finally

\be
\begin{array}{l}
\langle \left[
  h\left(\vec{x},0\right)-h\left(\vec{x'},0\right)\right]^2\rangle=\\
\frac{k_BT}{4\pi}\left|\vec{x}-\vec{x'}\right|^2\left\{\ln\left(\frac{L}{\left|\vec{x}-\vec{x'}\right|}\right)
+0.5-\frac{1}{4\pi}\int_{\left|\vec{x}-\vec{x'}\right|/L}^{1}\frac{dy}{y^3}\times\right.\\
\left.
\times\left(4\pi
  y^2+H_{2\,4}^{2\,2}\left[y\left|{\begin{array}{cccc}
\left(0,1\right)&\left(\frac{5}{6},\frac{1}{6}\right)& \\
\left(0,\frac{1}{2}\right)&\left(\frac{5}{6},\frac{1}{6}\right)&\left(0,\frac{1}{2}\right)&\left(1,1\right)\\
\end{array} } \right.\right.\right]\right)-\\
\left.-\frac{1}{4\pi}\int_{0}^{1}dy H_{4\,2}^{2\,2}\left[y\left|{\begin{array}{cccc}
\left(\frac{3}{2},\frac{1}{2}\right)&\left(\frac{1}{3},\frac{1}{6}\right)&\left(\frac{3}{2},\frac{1}{2}\right)&\left(1,1\right) \\
\left(2,1\right)&\left(1,\frac{5}{6}\right)\\
\end{array} } \right.\right]\right\}.
\label{app:corr_ZG_3}
\end{array}
\ee

\noi Expanding  the $H$-functions in the former expressions by
use of (\ref{exp}) we can then get the corrections to
 0.5 in the brackets.

%%%%%%%%%%%%%%%%%%%%%%%%%%%%%%%%%%%%%%%%%%%%%%%%%%%%%%%%%%%%%%%%%%%%%%%%%%%%%%%
%
%                      A P P E N D I X E
%
%%%%%%%%%%%%%%%%%%%%%%%%%%%%%%%%%%%%%%%%%%%%%%%%%%%%%%%%%%%%%%%%%%%%%%%%%%%%%%%
\section{Generalized Langevin equation for $\Delta_{DA}(t)$}
\label{app:D-A_GLE}

In this appendix we derive the generalized Langevin equation (GLE) for the single component of the donor-acceptor distance $\Delta_{D-A}(t)$ in a protein, and then evaluate its autocorrelation function as arising from the outlined framework. The derivation of the GLE, as well as the corresponding correlation function, traces that proposed in  ~\cite{Lizana} for single file systems: the main difference is that in our analysis we will make  use  of the Fourier transform instead of Laplace in the time domain.

\noi Subtracting the FLE (\ref{FLE}) for the donor position  $h(x_D,t)$ from the FLE for the acceptor $h(x_A,t)$, the corresponding FLE for $\Delta_{D-A}(t)$ is achieved:

\be
2D_+^{1/2}\Delta_{D-A}(t)=\zeta_{D-A}(t)
\label{FLE_DA}
\ee

\noi where we implicitly assumed that $x_D<x_A$ along the protein backbone, and  $\zeta_{D-A}(t)=\zeta(x_A,t)-\zeta(x_D,t)$.
Eq.(\ref{FLE_DA}) does not satisfy the generalized FD relation, as it is straightforwardly shown by calculating the correlation function of the noise

\be
\begin{array}{l}
\langle\zeta_{D-A}(t)\zeta_{D-A}(t')\rangle=\\
2\left(\langle\zeta(x_A,t)\zeta(x_A,t')\rangle-\langle\zeta(x_A,t)\zeta(x_D,t')\rangle\right).
\label{noise_corrfunc_DA}
\end{array}
\ee

\noi The second term in the  RHS of the former expression can be derived using the general formula (\ref{FGN_cc_3}) or by direct calculation, once one notices that in this simple case the function $\Phi(x,\omega)$ appearing in (\ref{FGN}) is

\be
\Phi(x,\omega)=e^{-\left|x\right|\sqrt{-i\omega}}
\label{Phi_DA}
\ee

\noi As a matter of fact, in the $\omega$ space the FD relation reads ~\cite{Kubo}

\be
\langle\zeta(\omega)\zeta(\omega')\rangle=4\pi k_BT \,\Re e\left[\gamma(\omega)\right]\delta(\omega+\omega')
\label{FD_F}
\ee

\noi where $\gamma(\omega)$ represents the Fourier transform of the damping kernel , which in (\ref{FLE_DA}) is given by

\be
\gamma_{D-A}(\omega)=\frac{2}{\sqrt{-i\omega}},
\label{gamma_DA}
\ee

\noi and, on the other hand,

\be
\begin{array}{l}
\langle\zeta_{D-A}(\omega)\zeta_{D-A}(\omega')\rangle=\\
\frac{8\pi k_BT\sqrt{2}}{\left|\omega\right|^{1/2}}\left[1-e^{-\chi_{D-A}(\omega)}\left(\cos\chi_{D-A}(\omega)-\sin\chi_{D-A}(\omega)\right)\right]
\end{array}
\label{noise_corre_func_F_DA}
\ee

\noi with $\chi_{D-A}(\omega)=\frac{\left|\omega\right|^{1/2}(x_A-x_D)}{\sqrt{2}}$.

\noi To restore the validity of the FD relation we multiply both the terms of
(\ref{FLE_DA}) by

\be
C(\omega)=\frac{1-e^{\sqrt{-i\omega}(x_a-x_D)}}{2\left(1+e^{-2\chi_{D-A}(\omega)}-2e^{-\chi_{D-A}(\omega)}\cos\chi_{D-A}(\omega)\right)}
\label{C_omega_DA}
\ee

\noi obtaining the following form of the GLE in the Fourier space

\be
\tilde{\gamma}_{D-A}(\omega)(-i\omega)\Delta_{D-A}(\omega)=\tilde{\zeta}_{D-A}(\omega)
\label{FLE_DA_1}
\ee

\noi with $\tilde{\gamma}_{D-A}(\omega)=\frac{1}{\sqrt{-i\omega}\left(1-e^{\sqrt{-i\omega}(x_a-x_D)}\right)}$ and $\tilde{\zeta}_{D-A}(\omega)=C(\omega)\zeta_{D-A}(\omega)$. Now, it is clear that for small $\omega$ the Fourier transform of the damping kernel gets to the asymptotic  value

\[
 \tilde{\gamma}_{D-A}(\omega)\to\frac{1}{(-i\omega)(x_A-x_D)}.
\]

\noi Therefore we sum and subtract on the RHS of (\ref{FLE_DA_1}) the asymptotic expression $\frac{1}{(-i\omega)(x_A-x_D)}$, obtaining

\be
\tilde{\gamma}^{eff}_{D-A}(\omega)(-i\omega)\Delta_{D-A}(\omega)=-\omega^2_0\Delta_{D-A}(\omega)+\tilde{\zeta}_{D-A}(\omega)
\label{FLE_DA_3}
\ee

\noi where $\tilde{\gamma}^{eff}_{D-A}(\omega)=\tilde{\gamma}_{D-A}(\omega)-\frac{1}{(-i\omega)(x_A-x_D)}$ and $\omega^2_0=\frac{1}{x_A-x_D}$.

\noi It is immediate to show that the FD relation (\ref{FD_F}) still holds and  inverting in  time domain we finally get the sought form of the GLE for the donor-acceptor distance which tends to an FLE with fractional derivative of order 1/2 in the long time limit ~\cite{Lizana,Taloni-FLE,Kou,Min}

\be
\int_{-\infty}^t\tilde{\gamma}^{eff}_{D-A}(t-t')\frac{d}{dt'}\Delta_{D-A}(\omega)=-\omega^2_0\Delta_{D-A}(t)+\tilde{\zeta}_{D-A}(t)
\label{FLE_DA_4}
\ee

\noi We now want to evaluate the donor-acceptor autocorrelation function: we firstly need the expression of  $\Delta_{D-A}(\omega)$ from (\ref{FLE_DA_3}), i.e.

\be
\Delta_{D-A}(\omega)=\frac{\tilde{\zeta}_{D-A}(\omega)\left(1-e^{\sqrt{-i\omega}(x_a-x_D)}\right)}{\sqrt{-i\omega}}
\label{sol_FLE_DA}
\ee

\noi from whom, thanks to the  noise autocorrelation function

\be
\begin{array}{l}
\langle\tilde{\zeta}_{D-A}(\omega)\tilde{\zeta}_{D-A}(\omega')\rangle=\\
\frac{2\pi k_BT\sqrt{2}}{\left|\omega\right|^{1/2}}\frac{1-e^{-\chi_{D-A}(\omega)}\left(\cos\chi_{D-A}(\omega)-\sin\chi_{D-A}(\omega)\right)}{1+e^{-2\chi_{D-A}(\omega)}-2e^{-\chi_{D-A}(\omega)}\cos\chi_{D-A}(\omega)},
\end{array}
\label{noise_corre_func_F_DA_1}
\ee

\noi we derive the correlation function
\be
\begin{array}{l}
\langle\Delta_{D-A}(\omega)\Delta_{D-A}(\omega')\rangle=\frac{2\pi k_BT\sqrt{2}}{\left|\omega\right|^{3/2}}\times\\
\times\left[1-e^{-\chi_{D-A}(\omega)}\left(\cos\chi_{D-A}(\omega)-\sin\chi_{D-A}(\omega)\right)\right]\delta(\omega+\omega')
\end{array}
\label{distance_corre_func_F_DA}
\ee

\noi Hence inverting in time the previous expression and changing variable ($y=\left|\omega\right|^{1/2}(x_A-x_D)/2$) we finally get

\be
\begin{array}{l}
\langle\Delta_{D-A}(t)\Delta_{D-A}(t')\rangle=\frac{\sqrt{2} k_BT}{\pi}(x_A-x_D)\times\\
\times\int_0^{\infty}dy\frac{\cos\left(\frac{4y^2\left|t-t'\right|}{(x_A-x_D)^2}\right)}{y^2}\left[1-e^{-\sqrt{2}y}\left(\cos\sqrt{2}y-\sin\sqrt{2}y\right)\right]
\end{array}
\label{distance_corre_func_DA}
\ee

\noi which  exactly matches the expression furnished in (\ref{D-A_corr_2}).


\begin{thebibliography}{99}
\bibitem{footnote} According to (\ref{EW_corr_func_z=d}) the original formulation of the Edward-Wilkinson model cannot be taken since for $z=2$
and $d=2$ (see ~\cite{Edwards}) the asymptotic diffusion is logarithmic as already stressed in ~\cite{Barabasi}, in this case we then assume $z=1$.
\end{thebibliography}
\end{document}